\input lanlmac
\def\href#1#2{{#2}}

\input epsf.tex

\overfullrule=0mm

\newcount\figno
\figno=0
\def\fig#1#2#3{
\par\begingroup\parindent=0pt\leftskip=1cm\rightskip=1cm\parindent=0pt
\baselineskip=11pt
\global\advance\figno by 1
\midinsert
\epsfxsize=#3
\centerline{\epsfbox{#2}}
\vskip 12pt
{\bf Fig.\ \the\figno:} #1\par
\endinsert\endgroup\par
}
\def\figlabel#1{\xdef#1{\the\figno}}
\def\encadremath#1{\vbox{\hrule\hbox{\vrule\kern8pt\vbox{\kern8pt
\hbox{$\displaystyle #1$}\kern8pt}
\kern8pt\vrule}\hrule}}

   \font\fivebf  =cmbx10  scaled 500 
   \font\sevenbf =cmbx10  scaled 700 
   \font\tenbf   =cmbx10             
   \font\fivemb  =cmmib10 scaled 500 
   \font\sevenmb =cmmib10 scaled 700 
   \font\tenmb   =cmmib10            
\def\boldmath{\textfont0=\tenbf           \scriptfont0=\sevenbf 
              \scriptscriptfont0=\fivebf  \textfont1=\tenmb
              \scriptfont1=\sevenmb       \scriptscriptfont1=\fivemb}
\magnification=\magstep1
\baselineskip=12pt
\hsize=6.3truein
\vsize=8.7truein
 at 8truept
 at 8truept
 at 10truept

\vbox{\hfill IPhT-t12/055}
\bigskip
\bigskip
\font\bigrm=cmr12 at 14pt
\centerline{\bigrm Loop models on random maps via nested loops:}
\smallskip
\centerline{\bigrm case of domain symmetry breaking}
\smallskip
\centerline{\bigrm and application to the Potts model}

\bigskip\bigskip

\centerline{G. Borot$^1$, J. Bouttier$^2$ and E. Guitter$^2$}
  \smallskip
  \centerline{$^1$ Section de Math\'ematiques}
  \centerline{Universit\'e de Gen\`eve}
  \centerline{2-4 rue du Li\`evre, Case postale 64, 1211 Gen\`eve 4, Suisse}
  \centerline{$^2$ Institut de Physique Th\'eorique}
  \centerline{CEA, IPhT, F-91191 Gif-sur-Yvette, France}
  \centerline{CNRS, URA 2306}
\centerline{\tt gaetan.borot@unige.ch}
\centerline{\tt jeremie.bouttier@cea.fr}
\centerline{\tt emmanuel.guitter@cea.fr}

  \bigskip

     \bigskip\bigskip

     \centerline{\bf Abstract}
     \smallskip
     {\narrower\noindent
We use the nested loop approach to investigate loop models on random planar maps where the domains delimited by the loops are given two alternating colors, which can be
assigned different local weights, hence allowing for an explicit ${\bf Z}_2$ domain symmetry breaking. Each loop receives a non local weight $n$, as well as a local bending
energy which controls loop turns. By a standard cluster construction that we review, the $Q = n^2$ Potts model on general random maps is mapped to a particular instance
of this problem with domain-non-symmetric weights. We derive in full generality a set of coupled functional relations for a pair of generating series which encode the
enumeration of loop configurations on maps with a boundary of a given color, and solve it by extending well-known complex analytic techniques. In the case where loops 
are fully-packed, we analyze in details the phase diagram of the model and derive exact equations for the position of its non-generic critical points. In particular, we underline 
that the critical Potts model on general random maps is not self-dual whenever $Q \neq 1$. In a model with domain-symmetric weights, we also show the possibility of a spontaneous
domain symmetry breaking driven by the bending energy.

\par}

    \bigskip

\nref\TutteCPM{W.T. Tutte, {\it A census of planar maps}, Canad. J. of Math.
{\bf 15} (1963) 249-271.}
\nref\BIPZ{E. Br\'ezin, C. Itzykson, G. Parisi and J.-B. Zuber, {\it Planar
diagrams}, Comm. Math. Phys. {\bf 59} (1978) 35-51.}
\nref\LGMBuz{See for instance: J.-F. Le Gall and G. Miermont, {\it
Scaling limits of random trees and planar maps}, Lecture notes of the
Clay Mathematical Institute Summer School, Buzios (2010),
arXiv:1101.4856, and references therein.}
\nref\DGZ{See for instance: P. Di Francesco, P. Ginsparg and
J. Zinn--Justin, {\it 2D gravity and random matrices},
Physics Reports {\bf 254} (1995) 1-131,	arXiv:hep-th/9306153, 
and references therein.}
\nref\Kost{I. Kostov, {\it $O(n)$ vector model on a planar random lattice:
spectrum of anomalous dimensions}, Mod. Phys. Lett. {\bf 4} (1989) 217-226.}
\nref\EK{B. Eynard and C. Kristjansen, {\it Exact solution of the $O(n)$
model on a random lattice}, Nucl. Phys. {\bf B455} (1995) 577-618, 
arXiv:hep-th/9506193.}
\nref\BBG{G. Borot, J. Bouttier and E. Guitter, {\it A recursive approach to the $O(n)$ model on random maps via nested loops}, J. Phys. A: Math. Theor. 45 (2012) 045002, arXiv:math-ph/1106.0153.}
\nref\LGM{J.-F. Le Gall and G. Miermont, {\it Scaling limits of random planar 
maps with large faces}, Ann. Probab. {\bf 39(1)} (2011) 1-69, arXiv:0907.3262 
[math.PR].}
\nref\BBGa{G. Borot, J. Bouttier and E. Guitter, {\it More on the $O(n)$ model on random maps via nested loops: loops with bending energy}, J. Phys. A, {\bf 45} (2012) 275206, arXiv:math-ph/1202.5521.}
\nref\GJSZ{A.~Guionnet, V.F.R.~Jones, D.~Shlyakhtenko and P.~Zinn-Justin, {\it Loop models, random matrices and planar algebras}, to appear in Comm. Math. Phys. (2012), arXiv:1012.0619 [math.OA].}
\nref\BBM{O. Bernardi and M. Bousquet-M\'elou, {\it Counting colored planar maps: algebraicity results}, J. Comb. Theory B {\bf 101}, 1 (2011) 315-377, arXiv:math.CO/0909.1695.}
\nref\FK{C.M. Fortuin and P.W. Kasteleyn, {\it On the random-cluster model. I- Introduction and relation to other models}, Physica {\bf 57} (1972) 536-564.}
\nref\Baxter{R.J. Baxter, S.B. Kelland and F.Y. Wu, {\it Equivalence of the Potts model or Whitney polynomial with an ice-type model}, J. Phys. A {\bf 9} (1976) 397-411.}
\nref\Nienhuis{B. Nienhuis, {\it Coulomb gas formulation of two-dimensional phase transitions}, in {\it Phase transition and critical phenomena}, eds. C. Domb and J.L. Lebowitz, {\bf 11} (1987).}
\nref\KazPotts{V. Kazakov, {\it Exactly solvable Potts models, bond- and tree-like percolation on dynamical (random) planar lattice}, Nucl. Phys. B (Proc. Suppl.) {\bf 4} (1988) 93-97.}
\nref\Daul{J.-M. Daul, {\it $Q$-states Potts model on a random planar lattice}, arXiv:hep-th/9502014.}
\nref\PottsZJ{P. Zinn-Justin, {\it The dilute Potts model on random surfaces}, J. Stat. Phys. {\bf 98} (2000) 245-264, arXiv:cond-mat/9903385.}
\nref\EynBon{B. Eynard and G. Bonnet, {\it The Potts-$q$ random matrix model : loop equations, critical exponents, and rational case}, Phys. Lett. {\bf B463} (1999) 273–279, arXiv:hep-th/9906130.}
\nref\NienDil{B. Nienhuis, {\it Analytical calculation of two leading exponents of the dilute Potts model}, J. Phys. A: Math. Gen. {\bf 15} (1982) 199-213.}
\nref\EZJ{B. Eynard and J. Zinn--Justin, {\it The $O(n)$ model on a random
surface: critical points and large order behaviour}, Nucl. Phys. {\bf B386}
(1992) 558-591, arXiv:hep-th/9204082.}
\nref\EKmore{B. Eynard and C. Kristjansen, {\it More on the exact solution
of the $O(n)$ model on a random lattice and an investigation of the
case $|n|>2$}, Nucl. Phys. {\bf B466} (1996) 463-487,
arXiv:hep-th/9512052.}
\nref\GBThese{G. Borot, Th\`ese de doctorat, Universit\'e d'Orsay (2011), arXiv:math-ph/1110.1493.}
\nref\KostSta{I.K. Kostov and M. Staudacher, {\it Multicritical phases of the 
$O(n)$ model on a random lattice}, Nucl. Phys. {\bf B384} (1992) 459-483,
arXiv:hep-th/9203030.}
\nref\KosGau{M. Gaudin and I.K. Kostov, {\it $O(n)$ model on a fluctuating planar lattice. Some exact results.}, Phys. Lett. B. {\bf 220} (1989), 200-206}
\nref\BoulK{D. Boulatov and V. Kazakov, {\it The Ising model
on a random planar lattice: the structure of the phase transition and the exact critical exponents}, Phys. Lett. B {\bf 186} (1987)
379-384.}
\nref\BMS{M. Bousquet-M\'elou and G. Schaeffer, {\it The degree distribution in bipartite planar maps: application to the Ising model}, preprint arXiv:math/0211070.}
\nref\Bicubic{J. Bouttier, P. Di Francesco and E. Guitter, {\it Combinatorics of bicubic maps with hard particles}, J. Phys. A: Math. Gen. {\bf 38} (2005), 4529-4559, arXiv:math/0501344.}
\nref\PZinnsixv{P. Zinn-Justin, {\it The six-vertex model on random lattices}, Europhys.Lett. {\bf 50} (2000) 15-21, arXiv:cond-mat.stat-mech/9909250.}
\nref\Kostovsixv{I.K.~Kostov, {\it Exact solution of the six-vertex model on a random lattice}, Nucl.Phys. B {\bf 575} (2000) 513-534, arXiv:hep-th/9911023.}
\nref\KostovADE{I.K.~Kostov, {\it Gauge invariant matrix model for the $\hat{A}-\hat{D}-\hat{E}$ closed strings}, Phys. Lett. B {\bf 297} (1992) 74-81, arXiv:hep-th/9208053.}
\nref\BEO{G.~Borot, B.~Eynard and N.~Orantin, work in progress.}

\newsec{Introduction}

The study of random planar maps, which are proper embeddings of graphs in the two-dimensional sphere, is
a fundamental issue in the fields of combinatorics \TutteCPM, as well as theoretical physics \BIPZ. When dealing with the limit of large maps,
many ensembles of random maps display the same asymptotic statistical properties, hence define the same {\it universality class}.
Such universality classes are characterized by a number of critical exponents, and in some cases, may be entirely determined in terms 
of a single, properly defined, probabilistic object.  
The simplest such object is the so-called Brownian map \LGMBuz, which, in the language of theoretical physics, characterizes the universality class of so-called
pure 2D quantum gravity and describes in particular the scaling limit of large
random maps with bounded face degrees. Other universality classes are reached when the maps have unbounded and
properly tuned face degrees, or when they are equipped with additional statistical models, driven to appropriate critical points \DGZ. 

A particularly important class of statistical models on random maps is formed by the $O(n)$ {\it loop models}, which consist 
in having maps endowed with configurations of self- and mutually-avoiding loops. In these models, a {\it non-local weight} $n$ is 
assigned to the loops, in addition to a number of local weights and, for $0\leq n\leq 2$,  the maps are known to display 
non-trivial scaling limits in several regions of parameters [\xref\Kost,\xref\EK].  
Phases of {\it dense loops} and {\it dilute loops} have been identified, each characterized by specific non-trivial exponents depending on $n$.

In \BBG, a new approach was introduced to study loop models, which consists in using the nested structure of the loops
to transform the model into a problem of maps without loops, but with faces of large degrees, controlled in a self-consistent way.
The advantages of the method is twofold: first, by establishing a direct relation between loop models and models of maps with controlled face degrees, it
 allows to identify the universality class of loop models in their dense and dilute phases with that of random maps with large face degrees
in their scaling limit. Such universality classes are entirely determined in terms of well-defined non-trivial probabilistic objects, themselves coded by stable
trees \LGM. The second advantage of the nested loop approach is that it provides an elegant
derivation of a number of functional equations for the generating functions of the loop model by simply expressing the self-consistency
of the mapping to maps with controlled face degrees. These equations were already known in some cases from a matrix integral formulation of the loop model at hand, and the nested loop approach gives them a clear combinatorial interpretation.
We applied this method in \BBGa\ to address a number of $O(n)$ loop models, incorporating in particular 
some {\it bending energy} for the loops.
Very explicit results were obtained for models where the loops visit only triangular faces, which extend the particular results of [\xref\Kost,\xref\EK]
for loops without bending energy.

So far, the method was applied only to loop models of the $O(n)$-type, which are ``domain-symmetric'' in the sense
that there is no parameter in the model that would introduce some distinction between the exterior and the interior of the loops.
The purpose of this paper is to study, on the contrary, loop models on random maps with {\it domain symmetry breaking}.
More precisely, let us define a {\it twofold loop configuration} on a map as a configuration of self- and mutually- avoiding loops
visiting the faces of the map, together with a choice of bicoloring, say in red and green, for the domains created
by the loop configuration. The color of domains is required to change upon crossing a loop so
that, in practice, the color of all domains is entirely fixed by that of a single one. All the vertices of the map, as well as 
those edges which are not crossed by loops, receive the color of the domain they lie in. The loops are nothing but
the domain walls for a ${\bf Z}_2$ variable which lives on the vertices and specifies their color. The notion of domain symmetry breaking is equivalent to that of {\it shading} for loop models studied in \GJSZ\ by random matrix techniques, and in the context of planar algebras.

In the $O(n)$-type models, red and green domains can always be introduced but, if so, they play symmetric roles, hence the color is an irrelevant variable. 
Here we shall be interested instead in situations where the red and green colors are treated non-symmetrically,  either because the parameters 
of the model are themselves not color-symmetric (explicit domain symmetry breaking), or because 
a spontaneous domain symmetry breaking occurs in some a priori color-symmetric model. As we shall explain below, 
the archetype of a loop model with domain symmetry breaking is obtained as the loop representation of the {\it Potts model on general random maps} (or, equivalently, the random-cluster model on general random maps). Besides, more general twofold loop models may be considered, including in particular 
some bending energy for the loops, which, as we shall see, may then be responsible for a spontaneous domain symmetry breaking in an initially symmetric model.
For completeness, let us mention that the Potts
model on general random maps was previously studied in \BBM\ by
different methods, while its corresponding loop model was also
considered in \GJSZ. While some of our results match those of this
latter paper, our approach does not rely on random matrix techniques,
and we furthermore discuss the critical behaviour of the model.

The paper is organized as follows: In Section 2, we introduce precisely the twofold loop models that we shall study. We show in particular
in Section~2.1 how to view the Potts model on general random maps as a particular instance of twofold loop model on random triangulations. Section~3 describes
the nested loop approach to these models via a gasket decomposition (Section~3.1), leading to coupled functional
equations for their generating series (Section~3.2). The solution of these equations is discussed in details in
Section~4 in a general setting, by introducing an appropriate elliptic parametrization. This framework is then used in Section~5 to address
the more specific question of non-generic critical points, where the maps display asymptotic behaviors which do not fall in the universality class of pure gravity. Explicit results are given in the case of the Potts model on random maps, as well as for fully-packed loop models with bending energy.
We show in particular in Section~5.5 how a spontaneous symmetry breaking may be generated by the bending energy in some 
a priori domain-symmetric model. We conclude in Section~6.

\newsec{Twofold loop models}

\subsec{The Potts model on random planar maps as a twofold loop model}

As it is well-known, the Potts model on a lattice may be expressed, very generally, as a model of loops [\xref\FK-\xref\Nienhuis], which turns out
to be a twofold loop model as we defined it. In this Section, we consider more precisely the Potts model defined on general random planar
maps and recall how to rephrase it as a twofold model of fully-packed loops on planar triangulations.

Given an arbitrary planar map ${\cal M}$, and for a given integer $Q\geq 1$, we define a Potts model on this map by assigning 
to each vertex $v$ of the map a Potts  variable $\sigma(v)$ taking its value in the set   $\{1,2,\ldots,Q\}$. Given such a Potts variable 
configuration, each edge $e$ of the map is then weighted by $t$ if the two Potts variables at its endpoints $v(e)$ and $v'(e)$ have
different values, and $t \exp(K)$ if they have the same value. In other words, if we denote by $E$, $V$ and $F$ the set of edges,
vertices and faces of ${\cal M}$, the weight of the Potts configuration $\{\sigma(v)\}_{v\in V}$ reads
\eqn\weightPotts{w(\{\sigma(v)\}_{v\in V})=t^{|E|}\prod_{e\in E} \exp{(K \delta_{\sigma(v(e))\sigma(v'(e))})},}
and the partition function $Z_{\rm Potts}({\cal M};t,K)$ of the Potts model on the map ${\cal M}$ is obtained by summing
over all Potts configurations, namely
\eqn\Zpotts{Z_{\rm Potts}({\cal M},t,K)\equiv \sum_{\{\sigma(v)\}_{v\in V}}   w(\{\sigma(v)\}_{v\in V}).}
We may eventually consider the grand-partition function 
\eqn\granfpf{{\hat Z}_{\rm Potts}(t,K;\mu_v,\mu_f)\equiv \sum_{{\rm{maps} \atop {\cal M}}}\mu_v^{|V|}\mu_f^{|F|}Z_{\rm Potts}({\cal M},t,v),}
where we also sum over all map configurations, with an additional weight $\mu_v$ per vertex
and $\mu_f$ per face of the map.

Writing $\exp{(K \delta_{\sigma\sigma'})}= 1+ J\delta_{\sigma\sigma'}$, with
\eqn\defv{J=\exp(K)-1,}
each factor in \weightPotts\ gives rise to two terms ($1$ or $J \delta_{\sigma(v(e))\sigma(v'(e))}$) and, upon expanding the product,
we may express the configuration weight as a sum over subsets $S$ of $E$ corresponding to those edges for which the term
$J \delta_{\sigma(v(e))\sigma(v'(e))}$ was chosen. This allows to recast the partition function of the Potts model on the map ${\cal M}$ as
\eqn\alterZpotts{Z_{\rm Potts}({\cal M},t,K = \ln(1 + J))= t^{|E|} \sum_{S\subseteq E} J^{|S|} Q^{c(S)},}
where the term $Q^{c(S)}$ arises from the sum over the Potts variables. Here $c(S)$ denotes the number of {\it clusters},
i.e.\ the number of connected components of the graph with vertex set $V$ and edge set $S$
(note that a cluster may be reduced to a single vertex). In this form, the model may be extended to arbitrary 
(non-necessarily integer) values of $Q$, and is then often called the {\it random-cluster model}.
The random-cluster model on random arbitrary planar maps may then be reformulated as a fully-packed twofold loop model on random planar triangulations as follows.

\fig{The equivalence between the random cluster model on arbitrary planar maps and the fully-packed twofold loop model on planar triangulations, see text.}{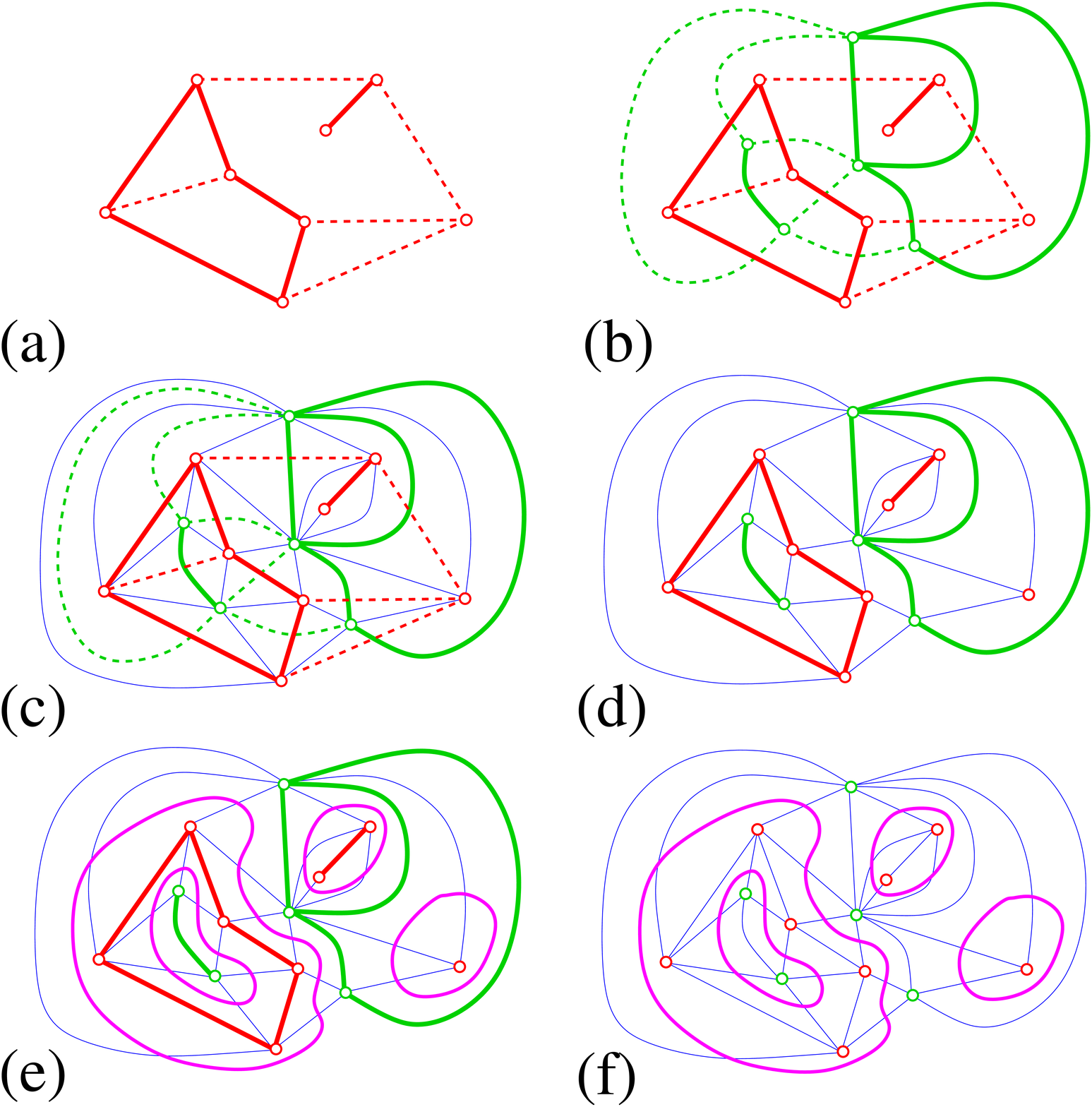}{12.cm}
\figlabel\pottstoloops
Given a map ${\cal M}$ endowed with a cluster configuration $S$, as displayed on Fig.~\pottstoloops-(a),
we introduce its dual map ${\cal M}^*$, with $V^*$ and $E^*$ its vertex and edge sets
($E^*$ being in bijection with $E$ and $V^*$ in bijection with $F$),
and we define $S^*$ as the subset of edges in $E^*$ whose dual edge is {\it not} in $S$.
In Fig.~\pottstoloops-(b), we represent 
simultaneously the maps ${\cal M}$ (red) and ${\cal M}^*$ (green), the edges in $S$ and $S^*$ being those displayed as solid lines.
Upon linking each vertex of ${\cal M}^*$ to all the vertices incident to its dual face in ${\cal M}$, we obtain a planar quadrangulation,
as displayed in blue on Fig.~\pottstoloops-(c), which is made of squares whose diagonals consist of an edge in $E$ and its dual edge in $E^*$. 
By definition of $S^*$, each square has exactly one of its diagonals in $S \cup S^*$. Thus, adding all the edges in $S \cup S^*$ to the quadrangulation, each square is split
into a pair of triangles (forming a ``diamond''), and we obtain a planar triangulation with vertex set $V \cup V^*$,
see Fig.~\pottstoloops-(d). By construction, each triangle is incident to exactly one edge in $S \cup S^*$, and we then draw an arch linking
the middles of the two other edges. As appears on Fig.~\pottstoloops-(e), those arches (displayed in purple) form a
configuration of closed self- and mutually-avoiding loops visiting the
faces of the triangulation. The loops visit {\it all the triangles} of
the triangulation, hence we say that the loop configuration is {\it
fully-packed}. Furthermore, the loop configuration is naturally a twofold loop configuration, 
as we defined it, with distinguished
red domains containing of all the vertices in $V$ and all the edges in $S$, and distinguished green domains, containing all 
the vertices in $V^*$ and all the edges in $S^*$. In Fig.~\pottstoloops-(f), we have suppressed the coloring of the edges since it is 
clearly a redundant information with the data of the loop configuration (and of the coloring of one vertex).  

Given conversely a planar triangulation endowed with a twofold fully-packed loop configuration, we may easily recover the planar map
endowed with a cluster configuration leading to it by following the above construction backwards. Each triangle being visited by a loop
has exactly one of its edge sides not visited by a loop. The other side of the corresponding edge necessarily belongs
to a different triangle and is the unique edge side of this triangle not visited by a loop. Concatenating the triangles by pairs
along their unvisited edge side therefore results into a planar quadrangulation, together with a marking of one diagonal for
each square. Now the bicoloring of the domains created by the twofold loop configuration on the triangulation 
induces a bicoloring in red and green of
the vertices of the quadrangulation (which is automatically bipartite). This allows to distinguish in each square of the quadrangulation
its red and green diagonals (colored according to the color of their endpoints), one of them being marked. The map ${\cal M}$ is then simply recovered by keeping only the red vertices and all the red (marked or not) diagonals, while the subset $S$ is simply the set of marked red diagonals. Note that exchanging 
the red and green colors in the twofold loop configuration leads to a different, and dual pre-image for the fully-packed configuration, namely $({\cal M}^*,S^*)$.

In conclusion, the above construction provides a one-to-one correspondence between planar maps with a distinguished edge subset
and triangulations endowed with a twofold fully-packed loop configuration.   
\medskip
Let us now see how to express the weight \alterZpotts\ in the language of the fully-packed loop configuration. Clearly, the
factor $t^{|E|}J^{|S|}$, which  corresponds to assigning a weight $Jt$ per edge in $S$ and a weight $t$ per edge in $E \setminus S$
in the original map, is recovered by assigning, on the triangulation, a weight $Jt$ per red edge (i.e.\ an edge strictly inside a red
domain) of the twofold loop 
configuration (edge in $S$), and a weight $t$ per green edge of the twofold configuration  (edge in $S^*$). 
Note that each triangle has its unvisited edge either red or green and we call it
red-facing and green-facing accordingly. The above weighting corresponds to assigning a weight $\sqrt{Jt}$ per red-facing triangle
(since an edge is shared by two triangles) and $\sqrt{t}$ per green-facing one. As for the weight $Q^{c(S)}$, it may be recovered as follows: a cluster (connected component) of $S$ made of ${\sl e}$ edges and ${\sl v}$ vertices yields a red domain which is surrounded by a number of loops equal to 
$2+{\sl e}-{\sl v}$. Indeed, it forms a planar map with ${\sl v}$ vertices, ${\sl e}$ edges 
and as many faces as surrounding loops so that the above expression is nothing but Euler's formula. Summing over all clusters, we deduce that the
total number of loops is $L=2c(S)+|S|-|V|$ since all vertices of the map ${\cal M}$ belong to a cluster by construction.  
The weight $Q^{c(S)}$ is then recovered by assigning a weight
\eqn\valn{n=\sqrt{Q}\quad \hbox{per loop},}
a weight $1/\sqrt{Q}$ per edge of $S$ (or equivalently $1/Q^{1/4}$ per red-facing triangle), and a weight $\sqrt{Q}$ per red vertex
in the triangulation. Gathering all the triangle weights, we find a weight
\eqn\triangleweights{\eqalign{h^{(1)} = \big(Jt/\sqrt{Q}\big)^{1/2}& \quad  \hbox{per red-facing triangle,}\cr
h^{(2)} = t^{1/2}\phantom{J\sqrt{Q}\,\,\,\,\,\,\,\,\,}& \hbox{per green-facing triangle.}\cr}}
As for the vertex weights in the triangulation, reintroducing the vertex and face weights $\mu_{v}$ and $\mu_f$ for ${\cal M}$, we get:
\eqn\vertexweights{\eqalign{u^{(1)} = \mu_{v}\,\sqrt{Q}\qquad & \quad \hbox{per red vertex,}\cr u^{(2)} = \mu_f\phantom{\sqrt{Q}}\qquad &  \hbox{per green vertex.}\cr}}
From the above discussion, we have the identity
\eqn\loopweight{\eqalign{{\hat Z}_{{\rm Potts}}(t,K;\mu_v,\mu_f)=\!\!\!\!\sum_{{\rm{triangulations}\atop {\cal T}}}\ \sum_{{\rm{twofold\ fully-packed\atop loop\ configs.\ on {\cal T}}}}\!\!\! \!\!\!\!\!\!\! n^{\#\rm{loops}}& (h^{(1)})^{\#{\rm{red-facing} \atop\rm{triangles}}}
(h^{(2)})^{\#{\rm{green-facing} \atop\rm{triangles}}}\cr & \times (u^{(1)})^{\#{\rm{red} \atop\rm{vertices}}}
(u^{(2)})^{\#{\rm{green} \atop\rm{vertices}}},\cr}}
with $K=\ln(1+J)$ and $n$, $h^{(1)}$, $h^{(2)}$, $u^{(1)}$ and $u^{(2)}$ as above.
Note that we may avoid introducing vertex weights in the triangulation (i.e. take $u^{(1)}=u^{(2)}=1$) by assigning, in the first place, a weight 
$\mu_v=1/\sqrt{Q}$ per vertex of the map ${\cal M}$, and no face weight ($\mu_f=1$), i.e.\ give the map ab initio a weight
\eqn\correctedweight{(\sqrt{Q})^{-|V|}Z_{\rm Potts}({\cal M},t,K).}
Note also that going from the Potts model on the map ${\cal M}$ to that on the dual map ${\cal M}^*$ simply amounts
to exchanging the red an green colors of the twofold loop configuration. We deduce the identity
\eqn\dualsym{(\sqrt{Q})^{-|V|}Z_{\rm Potts}({\cal M},t,K = \ln(1 + J))=(\sqrt{Q})^{-|V^*|}Z_{\rm Potts}({\cal M}^*,t^*,K^* = \ln(1+ J^*)),}
provided we choose $t^*$ and $J^*$ such that $t^*=Jt/\sqrt{Q}$ and $t=J^*t^*/\sqrt{Q}$ (that is to say $(h^{(2)})^*=h^{(1)}$ and 
$(h^{(1)})^*=h^{(2)}$), or equivalently
\eqn\tvtvstar{JJ^*=Q, \quad Jt^2=J^*t^{*2}.}
In this setting, self-duality is reached upon choosing $J=J^*=\sqrt{Q}$ and $t=t^*$.

\medskip

In conclusion, the Potts model on general random maps is equivalent to a twofold fully-packed loop model on random
triangulations, with weights \valn-\vertexweights. For the natural choice $\mu_v = \mu_f = 1$, we note that the vertex weights 
$u^{(1)}$ and $u^{(2)}$ are {\it not} color-symmetric, i.e. the model presents an {\it explicit domain symmetry breaking}. 
As we shall see, this implies that the triangles weights $h^{(1)}$ and $h^{(2)}$ are also, in general, not identical
at a critical point of the Potts models and their value cannot be obtained by a simple argument of self-duality.

\medskip

Let us insist on the fact that the correspondence holds between the set of {\it all} planar maps ${\cal M}$ carrying a Potts model on their vertices, and the set of all planar triangulations ${\cal T}$ carrying a twofold fully-packed loop model on their faces. We have not found a natural way to assign weights in ${\cal T}$ which would allow to control precisely the degree of the faces in ${\cal M}$. This prevents up to now the application of our method to the Potts model defined on, say, (the vertices of) random triangulations only. Such a Potts model has been studied using matrix integrals [\xref\KazPotts-\xref\EynBon] and more recently by combinatorial techniques \BBM.

\subsec{More general twofold loop models}

 \fig{Bending energy: a curvature weight $a$ is assigned to each pair of consecutive triangles facing the same domain (i.e. two red-facing or two green-facing triangles).}{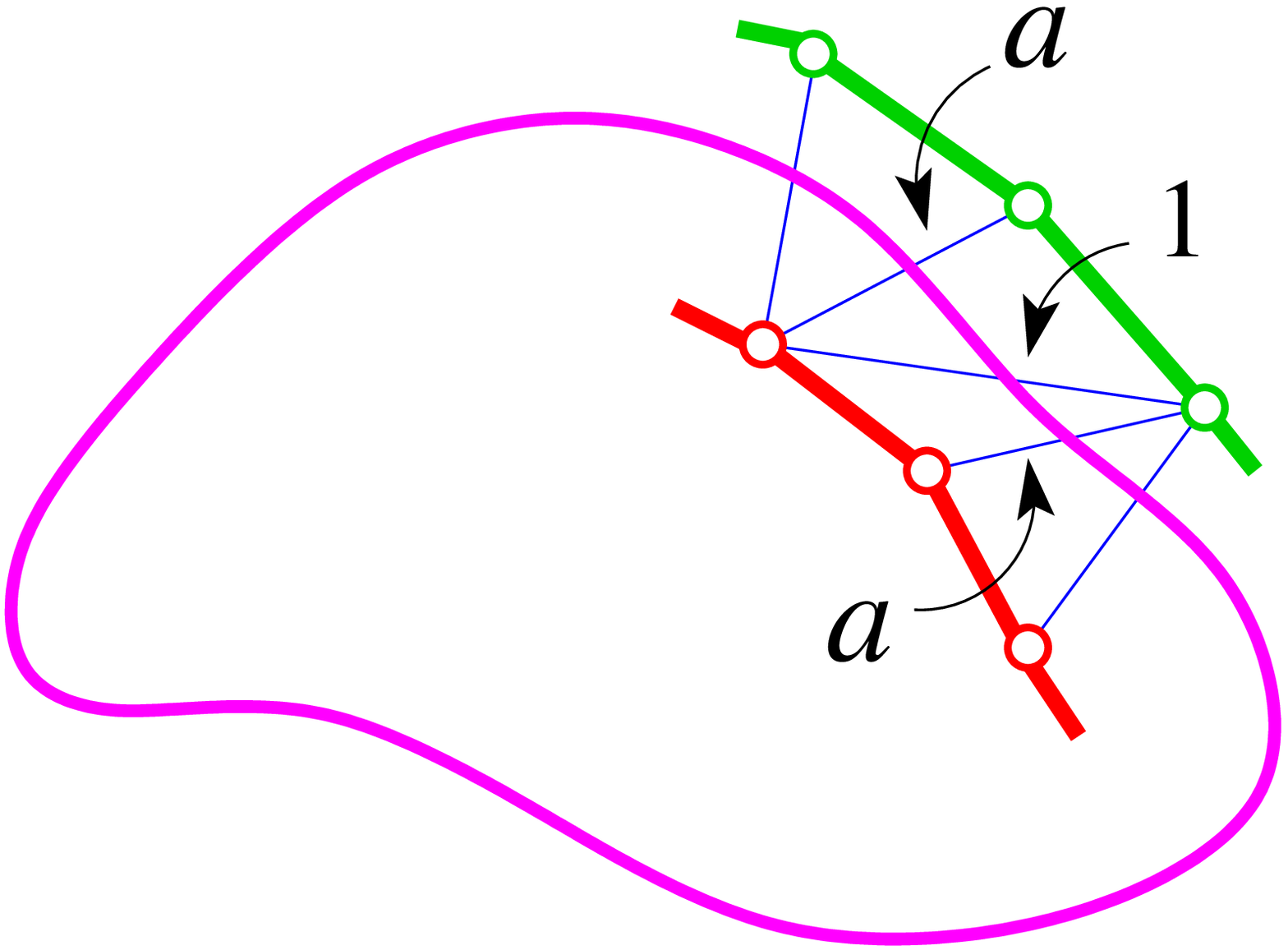}{6.cm}
 \figlabel\curvature

From the above discussion, the Potts model may be viewed as the archetype of twofold loop model with domain
symmetry breaking, and a particular attention will be paid to this model in the following. Nevertheless, we may want to consider more 
general twofold loop models with domain symmetry breaking. The main generalization discussed in this paper 
consists in introducing, as in \BBGa, some {\it bending energy} for the loops as follows: the triangles visited by a
given loop form a cyclic sequence of red- or green-facing triangles. To each pair of successive triangles along this
sequence, we attach a {\it curvature weight} $a$ if these triangles are the same nature (i.e. two red-facing triangles or two
green-facing ones) and a weight $1$ otherwise (i.e. one red-facing triangle and the other green-facing), see Fig.~\curvature.
The curvature weight has no obvious interpretation in the Potts model,
however in the random-cluster model \alterZpotts, it corresponds to a
corner interaction: to each corner of the map, we attach a weight $a$
if its two adjacent edges are both in $S$ or both in $E \setminus S$.
  \fig{Reformulation of the bending energy for twofold loop models on triangulations, as an Ising coupling on tetravalent maps (without loops).}{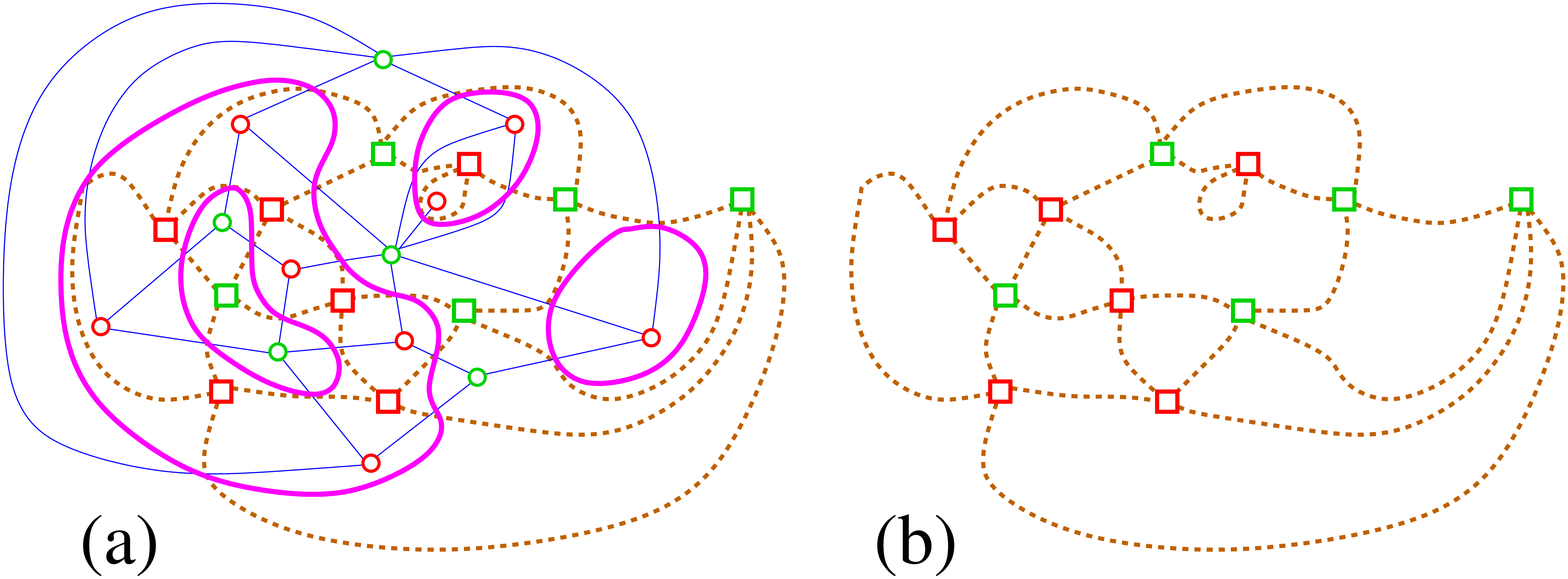}{14.cm}
\figlabel\isingcoupling
 \fig{A pair of consecutive triangles on a loop yields an edge in the tetravalent map. In case (a), both triangles are green-facing hence contribute a curvature weight $a$. In case (b), the triangles are of different nature hence no curvature weight is attached. This corresponds to attaching a weight $a$ to each monochromatic edge in the tetravalent map.}{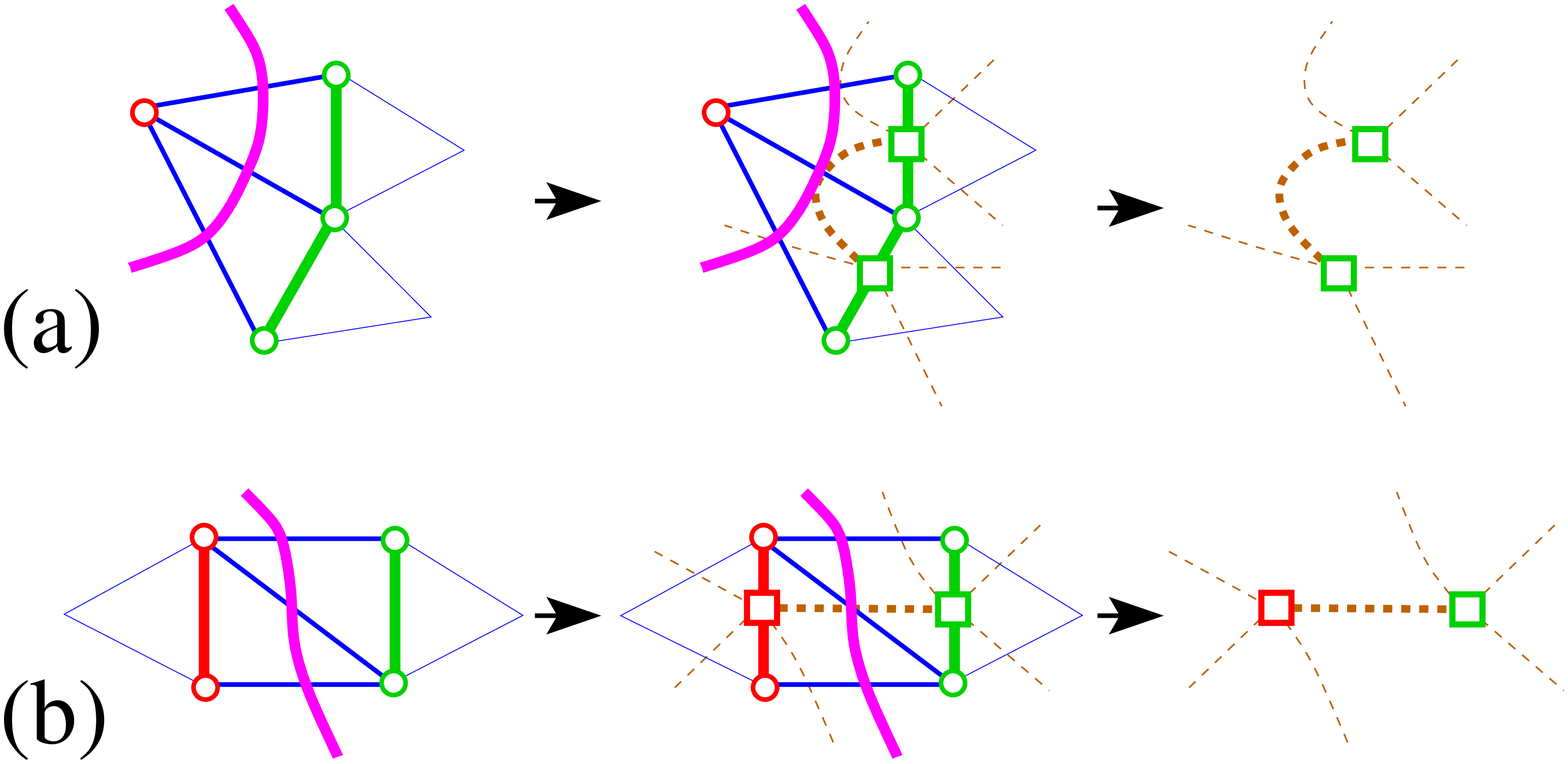}{14.cm}
 \figlabel\isingcouplingdetail
Alternatively, the curvature weight $a$ may be reformulated as an Ising-like coupling for spin variables $\{\varsigma\}$ as follows.
 We have seen that concatenating the triangles by pairs along their unvisited edge side  results into a planar quadrangulation with a marked, red or green diagonal in each square. Let us suppress this diagonal and replace it by a vertex at the center of the square,
with the same color, and consider the dual map of the quadrangulation, which is a tetravalent map (see Fig.~\isingcoupling).
The color of the square centers defines an Ising variable (say $\varsigma=+1$ for red and $\varsigma=-1$ for green) at the 
vertices of the dual tetravalent map. Now the loops cross all the edges of the quadrangulation. If we now
follow the cyclic sequence of triangles visited by a loop, we immediately see that having two consecutive triangles of
the same (red- or green-facing) nature corresponds to crossing an edge on the quadrangulation in such a way that the loop
remains tangent to the dual edge on the tetravalent map so that this dual edge connects two vertices carrying the
same Ising variable, see Fig.~\isingcouplingdetail-(a). Conversely,  having two consecutive triangles of
different nature corresponds to crossing an edge on the quadrangulation in such a way that the loop
also crosses the dual edge on the tetravalent map so that this dual edge now connects two vertices carrying different Ising variable, see Fig.~\isingcouplingdetail-(b). 
The curvature weight $a$ is equivalent to an Ising-like coupling of the form $1+(a-1) \delta_{\varsigma \varsigma'}$ for all pairs of adjacent
vertices on the dual tetravalent map. As for the the weights $h^{(1)}$ and $h^{(2)}$, they may be thought of as a magnetic-like
coupling of the form $(h^{(1)})^2\delta_{\varsigma,1}+(h^{(2)})^2\delta_{\varsigma,-1}$ for all vertices of the dual tetravalent map.  In particular, for $n=1$
and $u^{(1)}=u^{(2)}$, the twofold fully-packed loop model on random triangulations with curvature weight is fully equivalent to 
an Ising model on random tetravalent maps. We shall use this result later in this paper.

\medskip

As a last generalization of the model, we may finally release the constraint that
the loop configuration be fully-packed by allowing for faces which are not visited by loops. The loops are still
viewed as domain walls separating red and green domains and all the vertices of the map, as well 
as all edges not crossed by a loop and all faces not visited by a loop may be colored in red or green accordingly.
We shall let these unvisited faces have arbitrary degree $k\geq 1$ and weight them by $g_k^{(1)}$ accordingly if they lie in
a red domain, or $g_k^{(2)}$ if they lie in a green domain, with $(g_k^{(1)})_{k\geq 1}$ and $(g_k^{(2)})_{k\geq 1}$
two sequences of non-negative real numbers. The fully-packed model is then recovered by setting $g_k^{(1)}=g_k^{(2)}=0$ 
for all $k$. We also assign as before a weight $u^{(1)}$ per red vertex and  $u^{(2)}$ per green one. As for the faces visited 
by the loops, we still demand that they be triangles, among which we again distinguish red-facing and green-facing triangles, 
according to the color of their only edge  which is not crossed by a loop. The red-facing and green-facing triangles receive 
respective weights $h^{(1)}$ and $h^{(2)}$ as before. Finally, each loop receives a weight $n$, as well as some bending energy,
as defined above, with curvature weight $a$.

In the language of the Potts model, unvisited faces are known to correspond to defects or ``dilutions''.
More precisely, let us consider the so-called dilute Potts model where some vertices are {\it vacant}, i.e. carry no Potts variable
at all. In consequence, the weight 
\weightPotts\ is modified by restricting the product to the set of edges $E'$ whose both ends belong to the set $V'$ of non-vacant 
vertices. Performing the cluster expansion as before, we find that in \alterZpotts\ the sum is restricted to the subsets $S$ of $E'$, and 
$c(S)$ should be understood as the number of connected component of the graph $(V',S)$. Going to the loop representation,
each vacant vertex of degree $k$ yields a trivial red domain reduced to a single red vertex surrounded by a loop of length $k$.
Removing this trivial domain and its loop from the triangulation, we obtain a green unvisited face with degree $k$.
By adapting the counting arguments leading to \valn-\vertexweights, we find that the total number of loops in the resulting map
is now $L=2c(S)+|S|-|V'|$, hence each loop should still be counted with a weight $n=\sqrt{Q}$, while the weights
\triangleweights-\vertexweights\ are unchanged provided that each green unvisited face of degree $k$ receives a weight
$g_k^{(2)}=t^{k/2} \tilde{\mu}_v$. Here $\tilde{\mu}_v$ is a weight per vacant vertex in the original map,
which we can control independently from the weight $\mu_v$ for non-vacant vertices.
As for red unvisited faces, they correspond in the Potts language to ``dual'' defects, namely special faces
around which all the incident vertices are forced to carry the same Potts variable,
see for instance \NienDil\ for a detailed discussion in the context of the regular square lattice.
Clearly, such a face of degree $k$ yields 
in the loop representation a trivial green domain reduced to single green vertex surrounded by a loop of length $k$.
Removing this trivial domain, we obtain a red unvisited face with degree $k$ which should be counted with a weight
$g_k^{(1)}=(Jt)^{k/2} Q^{(2-k)/4} \tilde{\mu}_f$, where $\tilde{\mu}_f$ is a weight per special face in the original map
(by convention, edges incident to a special face are weighted $Jt$ instead of $(J+1)t$).
We emphasize that, since we are dealing with
the Potts model on arbitrary planar maps, the degrees of the unvisited faces of both colors
are not restricted. While the following techniques can be straightforwardly extended to this case to obtain a parametric answer, writing down as explicitly as possible the dependence in the initial parameters for the dilute Potts model is beyond the scope of this
paper. The dilute Potts model on {\it trivalent} maps was studied in \PottsZJ\ using matrix integral techniques but,
as mentioned at the end of the previous section, it is unclear how make the connection with our formalism by
translating the control on the vertex degrees in the loop model.

\medskip

In summary, our generalized twofold loop model with curvature weight
 depends on the following parameters: the vertex weights $u^{(1)}$,
 $u^{(2)}$, the visited triangle weights $h^{(1)}$, $h^{(2)}$, the
 unvisited face weights $(g_k^{(1)})_{k\geq 1}$ and
 $(g_k^{(2)})_{k\geq 1}$, the curvature weight $a$ and the weight $n$
 per loop.

\subsec{Maps with boundaries} 

Following the classical map enumeration methodology, we consider the
twofold loop model on maps with a boundary of length $\ell$, i.e.\
rooted maps (maps with a marked oriented edge) whose root face (the
face on the left of the root edge) has degree $\ell$. By convention,
we assume that the root face is not visited by loops (note that this
is always the case for $\ell \neq 3$, since the loops only visit
triangles in our model). This implies that the root face and all its
incident edges and vertices belong to a single colored domain, and we
say that the boundary is red or green accordingly. For $\ell \geq 1$, let us
denote by $F_\ell^{(1)}$ (respectively $F_{\ell}^{(2)}$) the generating function for maps with a twofold
loop configuration and with a red (respectively green) boundary of length $\ell$, where
for convenience we decide to assign to the root face a weight $1$ (instead of $g_\ell^{(1)}$ or $g_\ell^{(2)}$).
We also set $F_0^{(1)}=u^{(1)}$ and $F_0^{(2)}=u^{(2)}$ for the ``vertex-map'' reduced to a single red or green vertex
respectively. Note that, for $\ell=2$, the bivalent root face may be collapsed into a single root edge, thus
$F_2^{(1)}$ and $F_2^{(2)}$ count twofold loop configurations on rooted maps (without boundary), where the superscript
indicates the color of the root edge.

Interpreting the quantities $F_\ell^{(1)}$ and $F_\ell^{(2)}$ as observables in the Potts model involves some subtleties. 
In view of the discussion of the previous subsection, a red boundary of length $\ell$ corresponds in the Potts model to a root
face of degree $\ell$ around which all vertices are forced to carry the same Potts variable, which is akin to imposing fixed 
boundary conditions. On the other hand, a green boundary of length $\ell$ corresponds to a vacant root vertex of degree $\ell$,
which is close (but not precisely equivalent) to a root face of degree $\ell$ with free boundary conditions. Note however
that the quantity $F_2^{(1)}+F_2^{(2)}$ is equal to the partition function of the Potts model on rooted maps.

In the following, we will obtain a set of equations determining the $F_\ell^{(i)}$, which will be conveniently
rewritten by gathering all these generating functions into the two so-called ``resolvents''
 \eqn\resolvants{W^{(1)}(x)=\sum_{\ell\geq 0} {F_\ell^{(1)}\over x^{\ell+1}}, \qquad W^{(2)}(x)=\sum_{\ell\geq 0} 
 {F_\ell^{(2)}\over x^{\ell+1}}.}

\newsec{The nested loop approach to twofold loop models}

We now wish to study the statistical properties of the models introduced in Section 2, and in particular derive the location and 
properties of their critical points. To this end, we shall use the method introduced in \BBG\ and developed in \BBGa, which consists,
via some bijective decomposition, in deriving a set of functional equations for the resolvents of the model, as defined in \resolvants, 
and then in solving these equations along particular (critical) varieties. 

\subsec{The gasket decomposition}

Let us consider a map with a general twofold loop model, as defined in Section 2.2 above, and with a red boundary 
of some length $\ell$. Following \BBG, we shall define its {\it gasket decomposition} as follows: for each loop, we may distinguish
its exterior, which is the domain separated by this loop which contains the boundary, and its interior, which is the complementary
domain. In particular, we define the {\it outer} (resp. {\it inner}) contour of the loop as the set of edges that are incident 
to a triangle visited by the loop, and lie in the exterior (resp. interior) domain. Note that all the edges of the outer contour
of a given loop have the same color while all the edges of the inner contour have the opposite color. In other words,
each loop has a red and a green contour. Moreover, for a map with a red boundary, the outer contour of 
any {\it outermost} loop is red and its inner contour green. The gasket decomposition consists in cutting the
map along both the outer and inner contours of all the outermost loops. This disconnects the map into:
\item{$\bullet$} its gasket, which is the map spanned by the edges which were exterior to all loops. Clearly, all the vertices of the gasket are red vertices, thus weighted by $u^{(1)}$. As for its faces, they are either {\it regular faces} (corresponding to faces
that laid outside all outermost loops) weighted by $g_k^{(1)}$ according to their degree $k$, and {\it holes}, each delimited by the outer
contour of some outermost loop. If this outer contour has length $k$ (with $k\geq 1$), then the corresponding hole has degree
$k$. 
\item{$\bullet$} for each outermost loop:
\itemitem{-} its {\it ring} formed by the triangles visited by the loop. This ring is made of a sequence of $k$ red-facing triangles 
and $k'$ green-facing ones if the outer and inner contours of the loop at hand have respective lengths $k$ and $k'$.
\itemitem{-} its {\it internal map} which is the map spanned by the edges which were interior to the loop at hand. This internal map is 
itself a map with a green boundary of length $k'$ (the length of the inner contour of the loop at hand) endowed with a twofold loop 
configuration formed by all the loops that laid inside the loop at hand. For $k'=0$, the internal map reduces to the (green) vertex-map.

\medskip

 As explained in \BBG, all these different components are rooted objects whose root may be canonically obtained from that
of the original map by some (somewhat irrelevant) prescription. In particular, rings are rooted on their outer contour, or equivalently
have a marked red-facing triangle: we shall denote by $A_{k,k'}^{(1 \rightarrow 2)}(h^{(1)},h^{(2)})$ ($k\geq 1,k'\geq 0$)
the generating function for rings formed of $k$ red-facing and $k'$ green-facing triangles
weighted by $h^{(1)}$ and $h^{(2)}$ respectively, and with a marked red-facing triangle. Clearly, to enumerate properly all maps with 
a red boundary, we simply have to assign to each hole of degree $k$ a weight
\eqn\holeweight{n\, \sum_{k'\geq 0} A_{k,k'}^{(1 \rightarrow 2)}(h^{(1)},h^{(2)}) F_{k'}^{(2)}}
accounting for all possible rings with fixed outer length $k$ and varying inner length $k'$ and all internal maps
with a green boundary of compatible length $k'$, together with a weight $n$ for the outermost loop that gave rise
to the hole at hand. 

\subsec{The fixed-point conditions and the functional equations}

We may use the above gasket decomposition to relate the generating function $F_\ell^{(1)}$ of maps with a twofold
 loop configuration and with a red boundary of length $\ell$ to the generating function for planar maps {\it without loops} 
 but with a control on the degree of their faces. More precisely, let us introduce the generating function
 ${\cal F}_\ell(u\,;\,g_1,g_2,\ldots)$ for maps (with no loops) with a boundary of length $\ell$ and weight $u$ per vertex and $g_k$
 per face of degree $k$, with $(g_k)_{k\geq 1}$ a sequence of non-negative real weights (we set ${\cal F}_0(u\,;\,g_1,g_2,\ldots)=u$ by convention).  Then, from the gasket decomposition, we immediately see that $F_\ell^{(1)}$ is obtained
 by taking  ${\cal F}_\ell(u;g_1,g_2,\ldots)$ at $u=u^{(1)}$ and at particular values of $g_k$, given implicitly by
 \eqn\subst{g_k=g_k^{(1)}+n\, \sum_{k'\geq 0} A_{k,k'}^{(1 \rightarrow 2)}(h^{(1)},h^{(2)}) F_{k'}^{(2)}.}
 Here the two terms come from regular faces of degree $k$ in the gasket and holes of degree $k$ respectively. 
 Clearly, we obtain symmetrically $F_\ell^{(2)}$ by taking  ${\cal F}_\ell(u;g_1,g_2,\ldots)$ now at $u=u^{(2)}$ and 
 at values of $g_k$ given by
 \eqn\substtwo{g_k=g_k^{(2)}+n\, \sum_{k'\geq 0} A_{k,k'}^{(2 \rightarrow 1)}(h^{(1)},h^{(2)}) F_{k'}^{(1)},}
where
\eqn\Aexch{A_{k,k'}^{(2 \rightarrow 1)}(h^{(1)},h^{(2)}) = A_{k,k'}^{(1 \rightarrow 2)}(h^{(2)},h^{(1)}).}
To summarize, we may write the two identifications
\eqn\FcalF{F_\ell^{(1)}={\cal F}_\ell(u^{(1)}\,;\,G_1^{(1)},G_2^{(1)},\ldots), \qquad F_\ell^{(2)}={\cal F}_\ell(u^{(2)}\,;\,G_1^{(2)},G_2^{(2)},\ldots),}
where $G_k^{(1)}$ and $G_k^{(2)}$ are functions of all the parameters of the loop model, 
implicitly determined by the ``fixed-point conditions''
\eqn\fixedpoint{\eqalign{G_k^{(1)}&=g_k^{(1)}+n\, \sum_{k'\geq 0} A_{k,k'}^{(1 \rightarrow 2)}(h^{(1)},h^{(2)}) {\cal F}_{k'}(u^{(2)};G_1^{(2)},G_2^{(2)},\ldots),\cr
G_k^{(2)}&=g_k^{(2)}+n\, \sum_{k'\geq 0} A_{k,k'}^{(2 \rightarrow 1)}(h^{(1)},h^{(2)}) {\cal F}_{k'}(u^{(1)};G_1^{(1)},G_2^{(1)},\ldots).\cr}}
It is well-known (see Section 6 in \BBGa\ for a proof) that the resolvent 
\eqn\resol{{\cal W}(x)=\sum_{k\geq 0}{{\cal F}_\ell(u;g_1,g_2,\ldots)\over x^{\ell+1}}}
may be analytically continued into a function which is holomorphic on the complex plane except on a real 
interval $[\gamma_{-},\gamma_{+}]$, with $\gamma_{+}\geq |\gamma_{-}|$, where it has a discontinuity.
The discontinuity of ${\cal W}(x)$ on this cut is given by the so-called ``spectral density''
\eqn\spectral{\rho(x)={{\cal W}(x-{\rm i}0)-{\cal W}(x+{\rm i}0)\over 2{\rm i}\pi},}
which must be positive on $]\gamma_{-},\gamma_{+}[$ and must vanish at $x=\gamma_{\pm}$.
Moreover, ${\cal W}(x)$ satisfies the fundamental relation
\eqn\fundrel{\forall x\in ]\gamma_{-},\gamma_{+}[,\qquad {\cal W}(x+{\rm i}0)+{\cal W}(x-{\rm i}0)=V'(x),}
where we have set
\eqn\Vprime{V'(x)=x-\sum_{k\geq 1} g_k x^{k-1}.}
Remarkably enough, ${\cal W}(x)$, as well as $\gamma_{-}$ and $\gamma_{+}$, are then entirely fixed by those analytic properties and the behavior
\eqn\Wlatrgex{{\cal W}(x) \sim {u\over x}}
at large $x$. 
From the identification \FcalF, we deduce that $W^{(1)}(x)$ (resp. $W^{(2)}(x)$) can be analytically continued
into a function holomorphic on the complex plane except on a real interval $[\gamma^{(1)}_-,\gamma^{(1)}_+]$
(resp. $[\gamma_{-}^{(2)},\gamma_{+}^{(2)}]$), 
with $\gamma_{+}^{(1)}\geq |\gamma_{-}^{(1)}|$ 
(resp. $\gamma_{+}^{(2)}\geq |\gamma_{-}^{(2)}|$). The function $W^{(1)}(x)$ satisfies, for $x\in ]\gamma_{-}^{(1)},\gamma_{+}^{(1)}[$,
\eqn\fundrelonetwo{W^{(1)}(x+{\rm i}0)+W^{(1)}(x-{\rm i}0) = (V^{(1)})'(x)- n 
\sum_{k \geq 1}\sum_{k'\geq 0} A_{k,k'}^{(1 \rightarrow 2)}(h^{(1)},h^{(2)}) F^{(2)}_{k'} x^{k-1},}
and the function $W^{(2)}(x)$ satisfies, for $x \in  ]\gamma_{-}^{(2)},\gamma_{+}^{(2)}[$,
\eqn\fundrelonetwoB{W^{(2)}(x+{\rm i}0)+W^{(2)}(x-{\rm i}0) = (V^{(2)})'(x)- n 
\sum_{k \geq 1}\sum_{k'\geq 0} A_{k,k'}^{(2 \rightarrow 1)}(h^{(1)},h^{(2)}) F^{(1)}_{k'} x^{k-1},}
where we have set
\eqn\Vprimei{(V^{(i)})'(x)=x-\sum_{k\geq 1} g_k^{(i)}x^{k-1}, \quad i=1,2.}
The double sum in \fundrelonetwo\ may be rewritten as a contour integral around the cut of $W^{(2)}$, namely
\eqn\contint{\sum_{k \geq 1}\sum_{k'\geq 0} A_{k,k'}^{(1 \rightarrow 2)}(h^{(1)},h^{(2)}) F^{(2)}_{k'} x^{k-1}={1\over2{\rm i}\pi\,  x}
\oint_{[\gamma^{(2)}_-,\gamma^{(2)}_+]} {\rm d}y\, A^{(1 \rightarrow 2)}(x,y;h^{(1)},h^{(2)}) W^{(2)}(y)}
upon introducing the grand-canonical ring partition function
\eqn\gcrpf{A^{(1 \rightarrow 2)}(x,y;h^{(1)},h^{(2)})=\sum_{k \geq 1}\sum_{k'\geq 0} A_{k,k'}^{(1 \rightarrow 2)}(h^{(1)},h^{(2)}) x^{k}y^{k'}.}
Note that $]\gamma^{(2)}_-,\gamma^{(2)}_+[$ must be included in the
disk of convergence of $y \mapsto A^{(1 \rightarrow 2)}(x,y;h^{(1)},h^{(2)})$ for all $x$ in $]\gamma^{(1)}_-,\gamma^{(1)}_+[$ in order for the double sum to converge.

The grand-canonical ring partition function may easily be computed along the same lines as in \BBGa, with the result
\eqn\valgcrpf{\eqalign{A^{(1 \rightarrow 2)}(x,y;h^{(1)},h^{(2)})&={a h^{(1)} x \over 1-a h^{(1)} x}+x{\partial\over \partial x} \left(-\ln\left(1-{h^{(1)} x\over 1- a h^{(1)} x}{h^{(2)} y\over1-a h^{(2)} y}\right)\right)\cr &= {x s_1'(x)\over y-s_1(x)}+{x s_1''(x)\over 2 s_1'(x)},}}
where we have set
\eqn\sone{s_1(x)={1-a h^{(1)} x \over  a h^{(2)}+(1-a^2) h^{(1)} h^{(2)} x}.}
In the first line of \valgcrpf, the first term corresponds to $k'=0$ in \gcrpf, in which case the ring is made of red-facing triangles 
only, and the second term corresponds to $k'>0$ in which case the sequence of red- and green facing triangles may be viewed
as an alternance 
of blocks made of red-facing triangles only and blocks made of green-facing triangles only. Taking the $-\ln(\cdot)$ generates cyclic such
sequences, and the $x \partial/\partial x$ operator amounts to rooting the outer contour of the ring, as required.
The expression in the second line of \valgcrpf\ displays both the pole of $y \mapsto A^{(1 \rightarrow 2)}(x,y;h^{(1)},h^{(2)})$ at
$y=s_1(x)$ and its limit $x s_1''(x)/(2s_1'(x))$ when $y\to \infty$.
In particular, its radius of convergence is $|s_1(x)|$, which implies that
$s_1(]\gamma^{(1)}_-,\gamma^{(1)}_+[)$ and $]\gamma^{(2)}_-,\gamma^{(2)}_+[$ must be disjoint in order for \contint\ to converge.
Upon exchanging $h^{(1)}$ and $h^{(2)}$, we find a similar expression
\eqn\Atwoone{A^{(2 \rightarrow 1)}(x,y;h^{(1)},h^{(2)})={x s_2'(x)\over y-s_2(x)}+{x s_2''(x)\over 2 s_2'(x)},}
with 
\eqn\stwo{s_2(x)={1-a h^{(2)} x \over  a h^{(1)}+(1-a^2) h^{(1)} h^{(2)} x}.}
Note that the mappings $x\mapsto s_1(x)$ and $x\mapsto s_2(x)$ are locally decreasing, and they are reciprocal
\eqn\sonetwo{s_1(s_2(x))=s_2(s_1(x))=x.}
The contour integral in \contint\ is easily evaluated by the residue theorem. Its integrand $A^{(1 \rightarrow 2)}(x,y;h^{(1)},h^{(2)})W^{(2)}(y)$
has a pole at $y=s_1(x)$ with residue $-x s_1'(x)W^{(2)}(s_1(x))$ and a residue at infinity equal to
$u^{(2)}\,x s_1'(x)/(2 s_1(x))$ since $W^{(2)}(y) \sim u^{(2)}/y$ at large $y$.
We may eventually rewrite Eqn.~\fundrelonetwo\ as 
\eqn\fundrelonetwobis{W^{(1)}(x+{\rm i}0)+W^{(1)}(x-{\rm i}0) -n\,s_1'(x)W^{(2)}(s_1(x)) = (V^{(1)})'(x) 
- n\,u^{(2)}  {s_1''(x)\over 2 s_1'(x)},}
and Eqn.~\fundrelonetwoB\ as
\eqn\fundrelonetwobisB{
W^{(2)}(x+{\rm i}0)+W^{(2)}(x-{\rm i}0)-n\,s_2'(x)W^{(1)}(s_2(x)) = (V^{(2)})'(x) - n\,u^{(1)}  {s_2''(x)\over 2 s_2'(x)}.}
Again $W^{(1)}(x)$ and $W^{(2)}(x)$, as well as $\gamma_{\pm}^{(1)}$ and $\gamma_{\pm}^{(2)}$ are entirely fixed by
their analytic properties, relations \fundrelonetwobis-\fundrelonetwobisB\ and by the behavior at large $x$
\eqn\Wlatrgex{W^{(1)}(x) \sim {u^{(1)}\over x}, \qquad W^{(2)}(x) \sim {u^{(2)}\over x}.}
As before, the spectral densities
\eqn\spectraonetwol{\rho^{(i)}(x)={W^{(i)}(x-{\rm i}0)-W^{(i)}(x+{\rm i}0)\over 2{\rm i}\pi}, \qquad i=1,2}
must be positive on $]\gamma_{-}^{(i)},\gamma_{+}^{(i)}[$ respectively, and vanish respectively at $x=\gamma_{\pm}^{(i)}$.

To conclude, let us mention that the functional equations \fundrelonetwobis\ and \fundrelonetwobisB\ match Eqn.~(27) of \GJSZ\ up to notations, in the particular case corresponding to the Potts model ($a=1$, $(V^{(i)})'(x)=x$).

\newsec{General solution of the functional equations}

In this section, we take the functional relations \fundrelonetwobis-\fundrelonetwobisB\ as starting points, and we solve them 
by adapting the technology of \BBGa, itself inspired from [\xref\EK,\xref\EZJ,\xref\EKmore,\xref\GBThese]. The solution consists of four steps: (i) a reduction to homogeneous linear functional equations, (ii) a reformulation of these equations using some elliptic parametrization, leading to (iii) an explicit solution depending on the position of the discontinuities, considered as free parameters. The latter are eventually fixed by matching the divergent behaviors so as to fulfill \Wlatrgex. This last step becomes rather technical if one wishes to obtain explicit expressions for the position of the discontinuities. We instead focus on the construction of the phase diagram of the model by giving 
(as analytically explicit as possible) the location of its critical varieties. This program is achieved in Section 5 in the case of fully-packed loops.

We assume that $0 < n < 2$, which is the most interesting range for applications. This includes the limit $n \rightarrow 0$ where a single loop survives, and which is different from the trivial case $n = 0$. Our analysis holds also for $-2 < n < 0$ but we do not discuss it here. When $|n| > 2$, only the last step (matching the divergent behaviors) differs. The cases $n = \pm 2$ are special and could be treated by adapting the remarks of \KostSta\ to our setting.

\subsec{Reduction to homogeneous linear functional equations}

We may rewrite Eqn.~\fundrelonetwobis\ in a more symmetric way upon defining:
\eqn\tildeWtwo{W^{(2s)}(x)=s_1'(x)W^{(2)}(s_1(x)), \qquad (V^{(2s)})'(x) = s_1'(x)\,(V^{(2)})'(s_1(x)).}
Then we have, for $x\in ]\gamma_{-}^{(1)},\gamma_{+}^{(1)}[$,
\eqn\fundrelonetwoter{W^{(1)}(x+{\rm i}0)+W^{(1)}(x-{\rm i}0) -n W^{(2s)}(x)  = (V^{(1)})'(x) 
- n\,u^{(2)}  {s_1''(x)\over 2 s_1'(x)}}
and, for $x\in s_2(]\gamma_{-}^{(2)},\gamma_{+}^{(2)}[)$,
\eqn\funrelonetwoterB{
W^{(2s)}(x+{\rm i}0)+W^{(2s)}(x-{\rm i}0)-n W^{(1)}(x) = (V^{(2s)})'(x)
+ n\,u^{(1)}  {s_1''(x)\over 2 s_1'(x)}.}
Here, we used the reciprocity of $s_1$ and $s_2$, which implies
\eqn\aide{s_1'(x)s_2'(s_1(x))=1,\qquad {s_1'(x)s_2''(s_1(x)) \over 2\,s_2'(s_1(x))} =-{s_1''(x) \over 2\,s_1'(x)}.}
The condition \Wlatrgex\ on $W^{(2)}$ translates into the asymptotic behavior
\eqn\Wtwosinf{W^{(2s)}(x) = {-u^{(2)}\over x-s_2(\infty)}+O(1), \qquad x\to s_2(\infty).}

Since $n \neq \pm 2$, we can write
\eqn\Wparthom{W^{(1)}(x)=W^{(1)}_{{\rm part}}(x)+\overline{W}^{(1)}(x), \qquad W^{(2s)}(x) = W^{(2s)}_{{\rm part}}(x) + \overline{W}^{(2)}(x),}
where $W^{(1)}_{{\rm part}}(x)$ and $W^{(2s)}_{{\rm part}}(x)$ are particular solutions of \fundrelonetwoter, namely
\eqn\Wonetwospart{\eqalign{
W^{(1)}_{\rm part}(x) & = {2(V^{(1)})'(x)+n (V^{(2s)})'(x)\over 4-n^2} + {n(nu^{(1)} - 2u^{(2)}) \over 4-n^2}\,{s_1''(x)\over 2 s_1'(x)}, \cr
W^{(2s)}_{\rm part}(x) & = {n(V^{(1)})'(x)+2 (V^{(2s)})'(x)\over 4-n^2}+ {n(2u^{(1)}-n u^{(2)})\over 4-n^2}\,{s_1''(x)\over 2 s_1'(x)}.}}
Eqns.~\fundrelonetwoter\ and \funrelonetwoterB\ translate into the relations
\eqn\Whomeq{\eqalign{\overline{W}^{(1)}(x+{\rm i}0)+\overline{W}^{(1)}(x-{\rm i}0) -n \overline{W}^{(2)}(x) & = 0, \quad x\in ]\gamma^{(1)}_-,\gamma^{(1)}_+[, \cr
\overline{W}^{(2)}(x+{\rm i}0)+\overline{W}^{(2)}(x-{\rm i}0)-n\overline{W}^{(1)}(x) & = 0, \quad x\in s_2(]\gamma^{(2)}_-,\gamma^{(2)}_+[),}}
while Eqns.~\Wlatrgex\ and  \Wtwosinf\ dictate the singularities of the $\overline{W}^{(i)}$'s outside their discontinuity locus. 
We distinguish two cases:  $a=1$ and $a \neq 1$.

\medskip

\noindent $\bullet\,\,\boldmath{a = 1.}$ In this case, $s_1$ is an affine map, so $(V^{(1)})'(x)$ and $(V^{(2s)})'(x)$ are both polynomials. Then, the $\overline{W}^{(i)}$'s have a (multiple) pole at $\infty$, and since $s_1'' \equiv 0$ we find that, for $x\to \infty$,
\eqn\behlargxaun{\eqalign{\overline{W}^{(1)}(x) & = -{2(V^{(1)})'(x)+n (V^{(2s)})'(x)\over 4-n^2} + {u^{(1)} \over x} + o\left({1\over x}\right), \cr
\overline{W}^{(2)}(x) & = -{n(V^{(1)})'(x)+2 (V^{(2s)})'(x)\over 4-n^2} + {u^{(2)} \over x} + o\left({1\over x}\right).}}

\noindent $\bullet\,\,\boldmath{a \neq 1.}$ The pole of $s_1$ is now located at the finite value
\eqn\eqnpolpotB{s_2(\infty) = {- a \over (1 - a^2)h^{(1)}} \neq \infty,}
so $(V^{(1)})'(x)$ is a polynomial and $(V^{(2s)})'(x)$ is a rational function with a pole at $s_2(\infty)$ (and no residue at $x = \infty$). Then, the $\overline{W}^{(i)}$'s have (multiple) poles at $\infty$ and $s_2(\infty)$, and since
\eqn\srel{{s_1''(x)\over 2 s_1'(x)} = {- 1 \over x - s_2(\infty)},}
we find that, when $x \rightarrow \infty$,
\eqn\largea{\eqalign{\overline{W}^{(1)}(x) & = -{2(V^{(1)})'(x) \over 4-n^2} + { 2(2u^{(1)} - nu^{(2)}) \over 4 - n^2}\,{1 \over x} + 
o\left({1\over x}\right), \cr
\overline{W}^{(2)}(x) & = -{n(V^{(1)})'(x) \over 4-n^2} + {n(2u^{(1)}-n u^{(2)}) \over 4-n^2}\,{1 \over x} + o\left({1\over x}\right),}}
while, for $x \rightarrow s_2(\infty)$
\eqn\otherpole{\eqalign{\overline{W}^{(1)}(x) & = - {n(V^{(2s)})'(x) \over 4-n^2} + {n(nu^{(1)} - 2u^{(2)}) \over 4 - n^2}\,{1 \over x - s_2(\infty)} + O(1), \cr 
\overline{W}^{(2)}(x) & = -{2(V^{(2s)})'(x)\over 4-n^2} + {2(nu^{(1)} - 2u^{(2)}) \over 4 - n^2}\,{1 \over x - s_2(\infty)} + O(1).}}

\subsec{Discontinuities}

We shall consider for the time being that the discontinuity edges $\gamma_{\pm}^{(i)}$ are given. They will have to be determined later in terms of the parameters of the model (see Section~4.6). As explained in Section~3.2, $]\gamma_{-}^{(1)},\gamma_{+}^{(1)}[$ and $s_2(]\gamma_{-}^{(2)},\gamma_{+}^{(2)}[)$ do not overlap. In particular, when the parameters are small enough, $[\gamma_{-}^{(1)},\gamma_{+}^{(1)}]$ and $[\gamma_{-}^{(2)},\gamma_{+}^{(2)}]$ are small neighborhoods of $0$ and, since $s_2(0) = 1/ (ah^{(1)}) > 0$, $s_2([\gamma_{-}^{(2)},\gamma_{+}^{(2)}])$ is a small interval away from $0$, and to the right of $[\gamma_{-}^{(1)},\gamma_{+}^{(1)}]$.
Since $s_2$ is locally decreasing, with a pole at
\eqn\polpot{s_1(\infty) = {-a \over (1 - a^2)h^{(2)}} \neq 0,}
we have more precisely, for small enough parameters,
\eqn\order{s_2(\gamma_{-}^{(2)}) > s_2(\gamma_{+}^{(2)}) > \gamma_{+}^{(1)}>0>
\gamma_{-}^{(1)}.}
When the parameters varies, $s_2([\gamma_{-}^{(2)},\gamma_{+}^{(2)}])$ may swallow
$\infty$ and $s_2(\gamma_{-}^{(2)})$ jumps to the interval $]-\infty,\gamma_{+}^{(1)}[$. It should become clear later that the solution is perfectly regular across this jump. 

In any case, the real line can be decomposed in four sets with non-overlapping interiors:
\eqn\linedef{{\bf R}\cup\{\infty\} = C^{(1)} \cup C^{(2)} \cup D^{(1)} \cup D^{(2)},}
where:
\eqn\linedefB{C^{(1)} = [\gamma_{-}^{(1)},\gamma_{+}^{(1)}],\quad C^{(2)} = s_2([\gamma_{-}^{(2)},\gamma_{+}^{(2)}]),\quad D^{(1)} = [\gamma_{+}^{(1)},s_2(\gamma_{+}^{(2)})],}
and $D^{(2)}$ is swept by following ${\bf R}\cup\{\infty\}$ in positive direction starting from $s_2(\gamma_{-}^{(2)})$ and arriving at $\gamma_{-}^{(1)}$.

\subsec{Elliptic parametrization}

\medskip
\fig{Mapping of the complex plane with two cuts onto a rectangle via Eqn.~(4.18).}{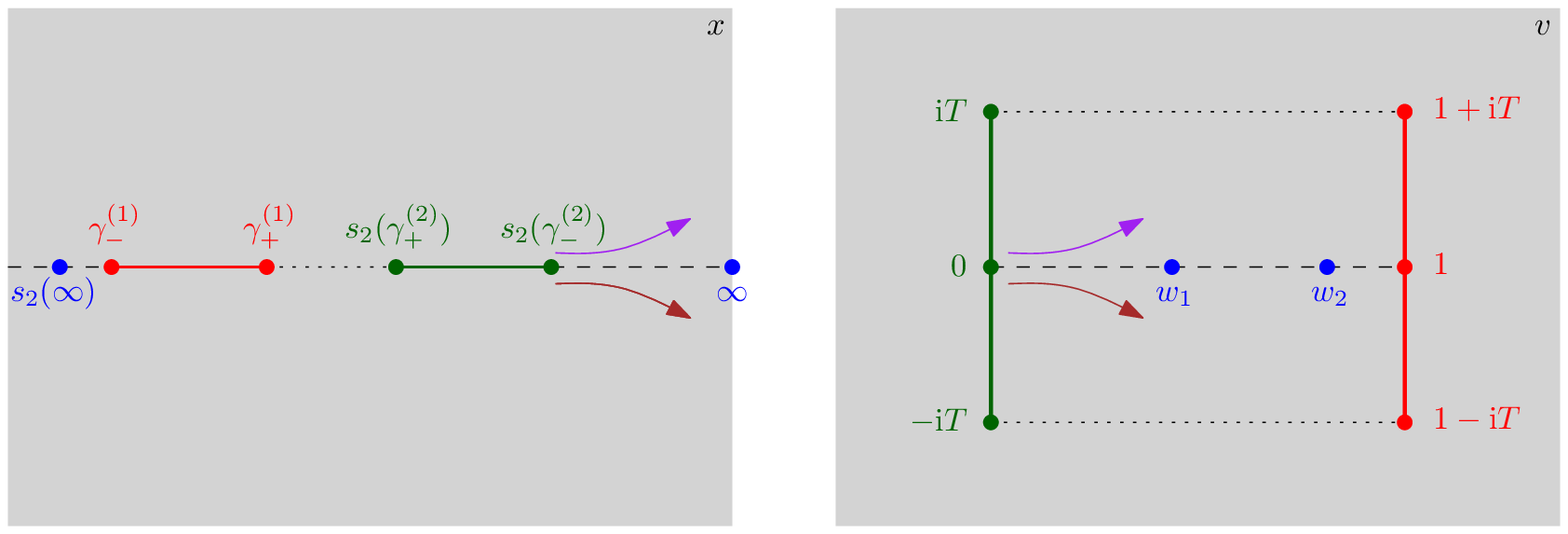}{13.cm}
\figlabel\ellipticparam

Our strategy is very similar to the one used in the context of $O(n)$ loop models with bending energy \BBGa, which correspond here to the special case $W^{(1)} \equiv W^{(2)}$. It is natural to introduce the elliptic integral which parametrizes the complex plane with two cuts $C^{(1)}$ and $C^{(2)}$, namely:
\eqn\ellpar{v(x) = c\,\int_{s_2(\gamma_{-}^{(2)})}^x {{\rm d}\xi \over \sqrt{\pm (\xi - \gamma_{-}^{(1)})(\xi - \gamma_{+}^{(1)})(\xi - s_2(\gamma_{-}^{(2)}))(\xi - s_2(\gamma_{+}^{(2)}))}}.}
We choose the sign $\pm$ so that the polynomial under the squareroot is non-negative on $D^{(2)}$, and a positive constant $c$ enforcing $v(D^{(2)}) = [0,1]$, see Fig.~\ellipticparam. Now, if we follow a path starting from $s_2(\gamma_{-}^{(2)})$ along $C^{(2)} + {\rm i}0$, $v$ runs over the positive imaginary axis, and we introduce $T \in\, ]0,+\infty]$ such that $v(C^{(2)} + {\rm i}0) = {\rm i}[0,T]$. Actually, $v$ maps the upper half-plane to $]0,1[ + {\rm i}]0,T[$, and the lower half-plane to $]0,1[ + {\rm i}]-T,0[$. Finally, we may check that $v$ maps $\gamma_{+}^{(1)}$ to $1 \mp {\rm i}T$. 
We introduce
\eqn\wdef{w_1 = v(+\infty + {\rm i}0),\qquad w_2 = v(s_2(+\infty + {\rm i}0)).}
If the local weights in our model are nonnegative, $w_1$ and $(1 - w_2)$ must belong to $]0,1[\cup{\rm i}]0,T[$. The inverse function $x(v)$ is a priori defined in the rectangle ${\cal R} = ]0,1[ + {\rm i}]-T,T[$, called the ``physical sheet''. By Schwarz reflection principle, it can be analytically continued to the whole complex plane as an even, doubly periodic function of periods $2$ and $2{\rm i}T$. Then,
\eqn\Vdef{\eqalign{{\cal V}^{(1)}(v) & = {{\rm d} V^{(1)}(x(v)) \over {\rm d} v}=x'(v)(V^{(1)})'(x(v)),\cr  \qquad {\cal V}^{(2)}(v) &= {{\rm d} (V^{(2)}\circ s_1)(x(v)) \over {\rm d} v}= x'(v)(V^{(2s)})'(x(v)),\cr}}
are defined as odd, doubly periodic meromorphic functions with periods $2$ and $2{\rm i}T$. Let us define similarly, {\it for $v$ in 
the physical sheet} ${\cal R}$,
\eqn\omegadef{\omega^{(i)}(v) = x'(v)\,\overline{W}^{(i)}(x(v)).}
We are going to extend these functions to the whole complex plane as follows.

\medskip

\item{$\bullet$} Since $(\overline{W}^{(i)}(x))_{i = 1,2}$ are continuous across $D^{(1)}$, we can extend $\omega^{(i)}$ as a meromorphic function on the strip ${\cal S} = ]0,1[ + {\rm i}{\bf R}$ by the relation
\eqn\extA{\omega^{(i)}(v + 2{\rm i}T) = \omega^{(i)}(v).}
\item{$\bullet$} Since $\overline{W}^{(1)}(x)$ is continuous across $C^{(2)}$, we can extend $\omega^{(1)}$ as a meromorphic function on the strip ${\cal S}^{(1)} = ]-1,1[ + {\rm i}{\bf R}$ by the relation
\eqn\extB{\omega^{(1)}(-v) = -\omega^{(1)}(v).}
\item{$\bullet$} Similarly, $\overline{W}^{(2)}(x)$ is continuous across $C^{(1)}$ so we can extend $\omega^{(2)}$ as a meromorphic function on the strip ${\cal S}^{(2)} = ]0,2[ + {\rm i}{\bf R}$ by the relation
\eqn\extC{\omega^{(2)}(2 - v) = -\omega^{(2)}(v).}
\item{$\bullet$} The behavior of $\overline{W}^{(1)}$ on its discontinuity (first line of Eqn.~\Whomeq) translates into
\eqn\extD{\forall v \in 1 + {\rm i}{\bf R} \qquad \omega^{(1)}(v) + \omega^{(1)}(v - 2) - n\omega^{(2)}(v) = 0,}
and this relation allows to extend the definition of $\omega^{(1)}$ to the strip $ ]1,2[ + {\rm i}{\bf R}$ and
via \extB\ to the strip $ ]-2,-1[ + {\rm i}{\bf R}$ so that  $\omega^{(1)}$ is now defined as a meromorphic function  on the larger strip 
$({\cal S}^{(1)})' = ]-2,2[ + {\rm i}{\bf R}$.
\item{$\bullet$} The behavior of $\overline{W}^{(2)}$ on its discontinuity (second line of Eqn.~\Whomeq) translates into
\eqn\extE{\forall v \in {\rm i}{\bf R}\qquad \omega^{(2)}(v) + \omega^{(2)}(v + 2) - n\omega^{(1)}(v) = 0,}
and this relation allows to extend the definition of $\omega^{(2)}$ to the strip $ ]-1,0[ + {\rm i}{\bf R}$, and
via \extC\ to the strip $ ]2,3[ + {\rm i}{\bf R}$ so that  $\omega^{(2)}$ is now defined as a meromorphic function on the larger strip $({\cal S}^{(2)})' = ]-1,3[ + {\rm i}{\bf R}$.
\item{$\bullet$} By recursion, we can then analytically continue $\omega^{(i)}$ as meromorphic functions on the whole complex plane which satisfy, for all $v \in {\bf C}$,
\eqn\extF{\eqalign{& \omega^{(1)}(v + 2{\rm i}T) = \omega^{(1)}(v), \cr & \omega^{(1)}(-v) = -\omega^{(1)}(v), \cr & \omega^{(1)}(v) + \omega^{(1)}(v - 2) - n\omega^{(2)}(v) = 0,}}
and
\eqn\extG{\eqalign{&\omega^{(2)}(v + 2{\rm i}T) = \omega^{(2)}(v), \cr & \omega^{(2)}(2 - v) = -\omega^{(2)}(v), \cr & \omega^{(2)}(v) + \omega^{(2)}(v + 2) - n\omega^{(1)}(v) = 0.}}
\item{$\bullet$} The spectral densities introduced in Eqn.~\spectraonetwol\ can be computed with the new parametrization
\eqn\densA{\eqalign{\rho^{(1)}(x(v + 1)) & = {\omega^{(1)}(v - 1) - \omega^{(1)}(v + 1) \over 2{\rm i}\pi\,x'(v + 1)},\quad v \in [-{\rm i}T,0], \cr 
\rho^{(2)}(s_1(x(v))) & = {\omega^{(2)}(v) - \omega^{(2)}(v + 2) \over 2{\rm i}\pi\,s'_1(x(v))\,x'(v)},\quad v \in [-{\rm i}T,0].}}

\subsec{Building the solution of Eqns.~\extF\ and \extG}

Let us denote by ${\bf T}$ the operator of translation by $1$ acting on the space of meromorphic, $2{\rm i}T$-periodic functions. The
last lines of \extF\ and \extG\ can be combined into
\eqn\reflG{({\bf T}^2 + n{\bf T} + 1)({\bf T}^2 - n{\bf T} + 1)\omega^{(i)} = 0, \qquad i = 1,2.}
Notice that, in the correspondence between the twofold loop model and the Potts model, we have $n = \sqrt{Q}$, and the equation above does not depend on the sign chosen for the square root. We parametrize the weight per loop as
\eqn\ndef{n = 2\cos\pi b,}
and the assumption $0 < n < 2$ is achieved by taking $0 < b < 1/2$. Then, we can decompose any solution of \reflG\ into a linear combination of pseudo-periodic functions, since the corresponding linear operator has the kernel decomposition
\eqn\deco{\eqalign{& \,\,{\rm Ker}({\bf T}^2 + n{\bf T} + 1)({\bf T}^2 - n{\bf T} + 1) \cr & \qquad  = {\rm Ker}({\bf T} - e^{{\rm i}\pi b})\,\oplus\,{\rm Ker}({\bf T} - e^{-{\rm i}\pi b})\,\oplus\,{\rm Ker}({\bf T} + e^{{\rm i}\pi b})\,\oplus\,{\rm Ker}({\bf T} + e^{-{\rm i}\pi b}).}}
There exists a unique function $v \mapsto \zeta_b(v)$ in $ {\rm Ker}({\bf T} - e^{{\rm i}\pi b})$ with a simple pole at $v = 0$ of residue $1$ and no other singularity modulo ${\bf Z} + {\rm i}T{\bf Z}$, i.e. satisfying
\eqn\polez{\zeta_b(v + 1) = e^{{\rm i}\pi b}\zeta_b(v),\qquad \zeta_b(v + 2{\rm i}T) = \zeta_b(v),\qquad \zeta_b(v) \mathop{\sim}_{v \rightarrow 0} {1 \over v}.}
It can be expressed as a ratio of Jacobi theta function as
\eqn\rsab{\zeta_b(v) = e^{{\rm i}\pi bv}\,{\vartheta_1(v + {\rm i}bT/2|{\rm i}T) \over \vartheta_1(v|{\rm i}T)}\,{\vartheta'_1(0|{\rm i}T) \over \vartheta_1({\rm i}bT/2|T)}.}
This function differs from that used in \BBGa\ by a modular transformation ${\rm i}T \rightarrow -1/({\rm i}T)$. Besides, Liouville theorem implies that the only bounded function in ${\rm Ker}({\bf T} - e^{{\rm i}\pi b})$ is zero, so any function in this subspace is uniquely characterized by the behavior at its poles in a fundamental domain, and can thus be expressed as a linear combination of derivatives $\partial_{w}^l\zeta_b(v - w)$, where $w$ denotes a pole in a fundamental domain, given that such a term yields a contribution
\eqn\polbeh{\partial_{w}^l\zeta_b(v - w) \mathop{=}_{v \rightarrow w} {l! \over (v - w)^{l + 1}} + O(1)}
to the divergent behavior. The description of the other subspaces in \deco\ is obtained by changing $(v,b)$ to 
$(-v,b)$, $(v,\tilde{b})$ and $(-v,\tilde{b})$ respectively, where $\tilde{b} = 1 - b$ is characterized by $-n = 2\cos\pi\tilde{b}$.

Now, we have to impose the second line of \extF, namely that $\omega^{(1)}$ be odd.
The general odd solution of \reflG\ with a simple pole at $w$ for some point $w$ in the physical sheet ${\cal R}$ takes the form
\eqn\ZdefA{\eqalign{Z^{(1)}(v;w,{\bf A}) & = A_-\big[\zeta_b(v - w) - \zeta_b(-v - w)\big] + A_+\big[\zeta_b(v + w) - \zeta_b(-v + w)\big] \cr 
& + \tilde{A}_-\big[\zeta_{\tilde{b}}(v - w) - \zeta_{\tilde{b}}(-v - w)\big] + \tilde{A}_+\big[\zeta_{\tilde{b}}(v + w) - \zeta_{\tilde{b}}(-v + w)\big]}}
with a vector of complex coefficients ${\bf A} = (A_-,A_+,\tilde{A}_-,\tilde{A}_+)$. Note that this solution also has a simple pole
at $1-w$, which also lies in the physical sheet. We may then generate
the general odd function with an $(l + 1)^{{\rm th}}$ order pole by
differentiating $l$ times $Z^{(1)}(v;w,{\bf A})$ with respect to
$w$. As we shall see in the next subsection, $\omega^{(1)}(v)$ will be
obtained by taking linear combination of such functions (with several
values of $w$, ${\bf A}$ and $l$). Note that, according to the third line of \extF, a term
$Z^{(1)}(v;w,{\bf A})$ in $\omega^{(1)}(v)$ gives rise in $\omega^{(2)}(v)$ to a term
\eqn\ZdefB{\eqalign{Z^{(2)}(v;w,{\bf A}) & = {1 \over n}\big[Z^{(1)}(v;w,{\bf A}) + Z^{(1)}(v - 2;w,{\bf A})\big] \cr
 & = A_-\big[e^{-{\rm i}\pi b}\zeta_b(v - w) - e^{{\rm i}\pi b}\zeta_b(-v - w)\big] + A_+\big[e^{-{\rm i}\pi b}\zeta_b(v + w) - e^{{\rm i}\pi b}\zeta_b(-v + w)\big] \cr 
& + \tilde{A}_-\big\{e^{{\rm i}\pi b}\zeta_{\tilde{b}}(v - w) - e^{-{\rm i}\pi b}\zeta_{\tilde{b}}(- v - w)\big] + \tilde{A}_+\big[e^{{\rm i}\pi b}\zeta_{\tilde{b}}(v + w) - e^{-{\rm i}\pi b}\zeta_{\tilde{b}}(-v + w)\big\} \cr}}
and those terms would precisely be canceled in \extG.
Moreover, those terms give rise in $\rho^{(1)}(v)\,x'(v + 1)$ to the term
\eqn\densB{\eqalign{\Delta^{(1)}(v;w,{\bf A}) & = {1 \over 2{\rm i}\pi}\big\{Z^{(1)}(v - 1;w,{\bf A}) - Z^{(1)}(v + 1;w,{\bf A})\big\}  \cr
& = {-\sin\pi b \over \pi}\Big\{A_-\big[\zeta_b(v - w) + \zeta_b(-v - w)\big] + A_+\big[\zeta_b(v + w) + \zeta_b(-v + w)\big] \cr 
& \phantom{{-\sin\pi b \over \pi}} + \tilde{A}_-\big[\zeta_{\tilde{b}}(v - w) + \zeta_{\tilde{b}}(-v - w)\big] + \tilde{A}_+\big[\zeta_{\tilde{b}}(v + w) + \zeta_{\tilde{b}}(-v + w)\big]\Big\}}}
and in $\rho^{(2)}(v)\,s_1'(x(v))x'(v)$ to the term
\eqn\densC{\eqalign{\Delta^{(2)}(v;w,{\bf A}) & = {1 \over 2{\rm i}\pi}\big\{Z^{(2)}(v;w,{\bf A}) - Z^{(2)}(v + 2;w,{\bf A})\big\} \cr
& = {-\sin\pi b \over \pi}\Big\{A_-\big[\zeta_b(v - w) + \zeta_b(-v - w)\big] + A_+\big[\zeta_b(v + w) + \zeta_b(-v + w)\big] \cr
& \phantom{{\sin\pi b \over \pi}} - \tilde{A}_-\big[\zeta_{\tilde{b}}(v - w) + \zeta_{\tilde{b}}(-v - w)\big] - \tilde{A}_+\big[\zeta_{\tilde{b}}(v + w) + \zeta_{\tilde{b}}(-v + w)\big]\Big\}.}}
Notice that $\Delta^{(2)}(v;w,{\bf A})$ differ from $\Delta^{(1)}(v;w,{\bf A})$ by a simple sign change, namely
\eqn\densing{\Delta^{(1)}(v;w,(A_-,A_+,\tilde{A}_-,\tilde{A}_+)) = \Delta^{(2)}(v;w,(A_-,A_+,-\tilde{A}_-,-\tilde{A}_+)).}

\subsec{Matching the divergent behavior}

We now describe a choice of basis functions which is appropriate for the type of divergent behavior encountered in our problem. For convenience, we distinguish the case $a = 1$ from the case $a \neq 1$.

\medskip

\noindent $\bullet\,\,\boldmath{a \neq 1.}$ For $i=1,2$, let us write the Laurent series of $\omega^{(i)}$ around $w_i$ as
\eqn\expL{\omega^{(i)}(v) \mathop{=}_{v \rightarrow w_i} \,\,\sum_{l \geq 0} {\alpha_{i|l} \over (v - w_i)^{l + 1}} + O(1),}
with $w_1$ and $w_2$ as in \wdef.
The coefficients $\alpha_{i|l}$ can be expressed from the first line of \largea\ (for $i=1$) and the second line of \otherpole\ (for $i=2$).
In particular, we have
\eqn\alphA{\alpha_{1|0} = -{2(2u^{(1)} - nu^{(2)}) \over 4 - n^2},\qquad \alpha_{2|0} = {2(nu^{(1)} - 2u^{(2)}) \over 4 - n^2},}
and we note that in the fully-packed case ($(V^{(i)})'(x)=x$), $\alpha_{i|l}$ vanishes for $l>2$.
Furthermore, by the second line of \largea\ and the first line of \otherpole, we have
\eqn\expM{\omega^{(3-i)}(v) \mathop{=}_{v \rightarrow w_{i}} \,\, {n \over 2} \sum_{l \geq 0} {\alpha_{i|l} \over (v - w_{i})^{l + 1}} + O(1).}
We must ensure that $\omega^{(i)}$ is regular at $v = 1 - w_1$ and $v = 1 - w_2$ for both $i=1,2$. This leads 
us to define vectors ${\bf A}_j \in {\bf C}^4$ for $j = 1,2$ as the unique solutions of the $4 \times 4$ systems of equations
\eqn\Xdefc{\eqalign{Z^{(1)}(v;w,{\bf A}_1) \mathop{\sim}_{v \rightarrow w} {1 \over v - w}, & \qquad Z^{(1)}(v;w,{\bf A}_1) \mathop{=}_{v \rightarrow 1-w} O(1), \cr
Z^{(2)}(v;w,{\bf A}_1) \mathop{\sim}_{v \rightarrow w} {n/2 \over v - w}, & \qquad Z^{(2)}(v;w,{\bf A}_1) \mathop{=}_{v \rightarrow 1 - w} O(1),}}
and
\eqn\Xdefd{\eqalign{Z^{(1)}(v;w,{\bf A}_2) \mathop{\sim}_{v \rightarrow w}  {n/2 \over v - w}, & \qquad Z^{(1)}(v;w,{\bf A}_2) \mathop{=}_{v \rightarrow 1 - w} O(1), \cr
Z^{(2)}(v;w,{\bf A}_2) \mathop{\sim}_{v \rightarrow w} {1 \over v - w}, & \qquad Z^{(2)}(v;w,{\bf A}_2) \mathop{=}_{v \rightarrow 1 - w} O(1).
}}
We find
\eqn\coefdef{{\bf A}_1 = {1 \over 4}\cdot(1,1,1,1),\qquad {\bf A}_2 = {1 \over 4}\cdot(e^{{\rm i}\pi b},e^{-{\rm i}\pi b},e^{-{\rm i}\pi b},e^{{\rm i}\pi b}).}
Now, if we define the two sets of functions ($j = 1,2$)
\eqn\Zde{\eqalign{Z_j^{(1)}(v) = Z_j^{(1)}(v;w_j,{\bf A}_j), & \qquad Z^{(2)}_j(v) = Z^{(2)}(v;w_j,{\bf A}_j), \cr 
\Delta^{(1)}_j(v) = \Delta^{(1)}(v;w_j,{\bf A}_j), & \qquad \Delta^{(2)}_j(v) = \Delta^{(2)}(v;w_j,{\bf A}_j).}}
we deduce that the desired solutions (matching all required behaviors) are
\eqn\omdef{\eqalign{\omega^{(1)}(v) & = {\cal D}_1\,Z_1^{(1)}(v) + {\cal D}_2\,Z_2^{(1)}(v), \cr \omega^{(2)}(v) & = {\cal D}_1\,Z_1^{(2)}(v) + {\cal D}_2\,Z_2^{(2)}(v),}}
where we introduce the differential operators
\eqn\diffop{{\cal D}_j = \sum_{l \geq 0} {\alpha_{j|l} \over l!}\,{\partial^l \over \partial w_j^{l}},\qquad j = 1,2.}
The corresponding spectral densities read
\eqn\densdef{\eqalign{
\rho^{(1)}(v) & = {{\cal D}_1\,\Delta_1^{(1)}(v) + {\cal D}_2\,\Delta_2^{(1)}(v) \over x'(v + 1)}, \cr
\rho^{(2)}(v) & = {{\cal D}_1\,\Delta_1^{(2)}(v) + {\cal D}_2\,\Delta_2^{(2)}(v) \over s'_1(x(v))\,x'(v)}.}}

\medskip

\noindent $\bullet\,\,\boldmath{a = 1.}$ Then, $w_1 = w_2$. The basis of functions we described for $a \neq 1$ is still well-defined and appropriate. Nevertheless, since $a = 1$ has an interest of its own, we prefer to define another basis of functions (related to the former one by a linear transformation), which takes Eqn.~\behlargxaun\ as a starting point: let us now write the Laurent series of $\omega^{(i)}$ around $w_1=w_2$ as
\eqn\expPP{\omega^{(i)}(v) \mathop{=}_{v \rightarrow w_1} \sum_{l \geq 0} {\beta_{i|l} \over (v - w_1)^{l + 1}} + O(1).}
In particular, we have
\eqn\alphB{\beta_{1|0} = -u^{(1)},\qquad \beta_{2|0} = -u^{(2)}.}
It leads us to define vectors ${\bf B}_j \in {\bf C}^4$ for $j = 1,2$ as the unique solutions of the $4 \times 4$ systems
\eqn\saudade{\eqalign{Z^{(1)}(v;w,{\bf B}_1) \mathop{\sim}_{v \rightarrow w} {1 \over v - w}, & \qquad Z^{(1)}(v;w,{\bf B}_1) \mathop{=}_{v \rightarrow 1 - w} O(1), \cr
Z^{(2)}(v;w,{\bf B}_1) \mathop{=}_{v \rightarrow w} O(1), & \qquad Z^{(2)}(v;w,{\bf B}_1) \mathop{=}_{v \rightarrow 1 - w} O(1),}}
and
\eqn\saudadf{\eqalign{Z^{(1)}(v;w,{\bf B}_2) \mathop{=}_{v \rightarrow w} O(1), \qquad & Z^{(1)}(v;w,{\bf B}_2) \mathop{=}_{v \rightarrow 1 - w} O(1), \cr Z^{(2)}(v;w,{\bf B}_2) \mathop{\sim}_{v \rightarrow w} {1 \over v - w}, \qquad & Z^{(2)}(v;w,{\bf B}_2) \mathop{=}_{v \rightarrow 1 - w} O(1).}}
We find
\eqn\Ydef{{\bf B}_1 = {1 \over 4{\rm i}\,\sin\pi b}\cdot(e^{{\rm i}\pi b},-e^{-{\rm i}\pi b},-e^{-{\rm i}\pi b},e^{{\rm i}\pi b}),\qquad {\bf B}_2 = {1 \over 4{\rm i}\,\sin\pi b}\cdot(-1,1,1,-1).}
If we now replace the definitions \Zde\ by
\eqn\YdefA{\eqalign{Z^{(1)}_j(v) = Z^{(1)}(v;w_1,{\bf B}_j),& \qquad Z^{(2)}_j(v) = Z^{(2)}(v;w_1,{\bf B}_j), \cr 
\Delta^{(1)}_j(v) = \Delta^{(1)}(v;w_1,{\bf B}_j), & \qquad \Delta^{(2)}_j(v) = \Delta^{(2)}(v;w_1,{\bf B}_j),}}
we deduce that the expressions~\omdef\ and \densdef\ for $\omega^{(i)}$ and $\rho^{(i)}$ are still valid 
provided we redefine the differential operators as
\eqn\diffredef{{\cal D}_j = \sum_{l \geq 0} {\beta_{j|l} \over l!}\,{\partial^l \over \partial w_1^l},\qquad j = 1,2.}

\subsec{Consistency conditions and position of the discontinuities}

We now come to the crucial question of the determination of the cuts in our problem. The following discussion 
is valid both in the case $a\neq 1$ with the expressions \Zde-\densdef, and in the case $a=1$, with the modified definitions \YdefA-\diffredef. 
The spectral density $\rho^{(1)}(x(v + 1))$ must vanish at $x(v+1) = \gamma^{(1)}_-$ (corresponding to $v = 0$) and at $x(v+1)= \gamma^{(1)}_+$ (corresponding to $v = -{\rm i}T$). Similarly, $\rho^{(2)}(s_1(x(v)))$ must vanish at $x(v) = s_2(\gamma^{(2)}_-)$ (corresponding to $v = 0$) and at $x(v) = s_2(\gamma^{(2)}_+))$ (corresponding to $v = -{\rm i}T$). This gives four independent conditions which determine in principle the positions of $\gamma^{(i)}_{\pm}$. 
By \densing, we can decompose $\Delta^{(i)}_j$ for $j = 1,2$ as
\eqn\decoBBB{\eqalign{\Delta^{(1)}_j(v) & = \Delta_{j}(v) + \tilde{\Delta}_{j}(v), \cr \Delta^{(2)}_{j}(v) & = \Delta_{j}(v) - \tilde{\Delta}_{j}(v),}}
where $\Delta_{j}$ only involves the function $\zeta_b$ and $\tilde{\Delta}_{j}$ only involves $\zeta_{\tilde{b}}$, see again
\densB\ and \densC. We deduce decoupled conditions
\eqn\POL{\eqalign{{\cal D}_1\Delta_{1}(0) + {\cal D}_2\Delta_{2}(0) & = 0,\cr
{\cal D}_1\Delta_{1}(-{\rm i}T) + {\cal D}_2\Delta_{2}(-{\rm i}T) & = 0,}}
and
\eqn\POLQ{\eqalign{{\cal D}_1\tilde{\Delta}_{1}(0) + {\cal D}_2\tilde{\Delta}_{2}(0) & = 0, \cr
{\cal D}_1\tilde{\Delta}_{1}(-{\rm i}T) + {\cal D}_2\tilde{\Delta}_{2}(-{\rm i}T) & = 0.}}

Last but not least, we must ensure that the spectral densities are positive in the interior of their support, which in general selects a unique solution to the above equations. Checking global positivity is a difficult task, that we do not carry otherwise than graphically in the examples of this article. However, one can often check analytically if the densities are locally positive at the edges of their support. We believe that this is actually sufficient:

\medskip

\noindent {\bf Conjecture.} Spectral densities for enumeration of maps in loop models are locally positive at the edges of their support iff they are globally positive on the whole support.

\medskip

\noindent This conjecture has been verified in all examples (maps with bounded face degree, maps with loops) we know of, and is also verified in the examples given later in this article.

\newsec{Non-generic critical points}

We shall now focus our discussion on the description of {\it non-generic critical points} (see \BBGa\ for a discussion on this terminology), corresponding to values of the parameters for which the generating functions for maps with a large boundary behave as
\eqn\largm{F_\ell^{(i)} \mathop{\sim}_{\ell \rightarrow \infty} C\,{(\gamma^{(i)}_+)^\ell \over \ell^{\nu +1}},}
with an exponent $\nu$ differing from the non-critical value $\nu=1/2$ and the generic critical value $\nu=3/2$.
By transfer, the resolvents $W^{(i)}(x)$ should display non-trivial singularities of the form $(\gamma^{(i)}_+ - x)^\nu$. This can only happen when
\eqn\overl{\gamma^{(1)}_+ = s_2(\gamma^{(2)}_+),}
i.e. when $T = \infty$ in the elliptic parametrization. Our main goal will be to derive the location of the non-generic critical variety in the parameters of the model. We carry this analysis as explicitly as possible in the case of fully-packed models, i.e. $g_k^{(i)} \equiv 0$ and thus $(V^{(1)})'(x) = (V^{(2)})'(x) = x$. The parameters of the model are then $u^{(1)},u^{(2)},h^{(1)},h^{(2)}$ the bending energy $a$ and the weight per loop $n$. We shall put emphasis on:

\medskip

\item{$\bullet$} the symmetric model: $u^{(1)} = u^{(2)}$ (and consequently $h^{(1)} = h^{(2)}$).
\item{$\bullet$} the Potts model in absence of additional vertex or face weights: $\mu_v = \mu_f = 1$, so that $u^{(1)}=\sqrt{Q}$ and $u^{(2)}=1$ according to Eqn.~\vertexweights.
\item{$\bullet$} the limit $n \rightarrow 0$ for general values of the parameters.

\subsec{Characterization}

\fig{The elliptic parametrization in the limit $\gamma^{(1)}_+ \rightarrow s_2(\gamma^{(2)}_+)$ via Eqn.~(5.4).}{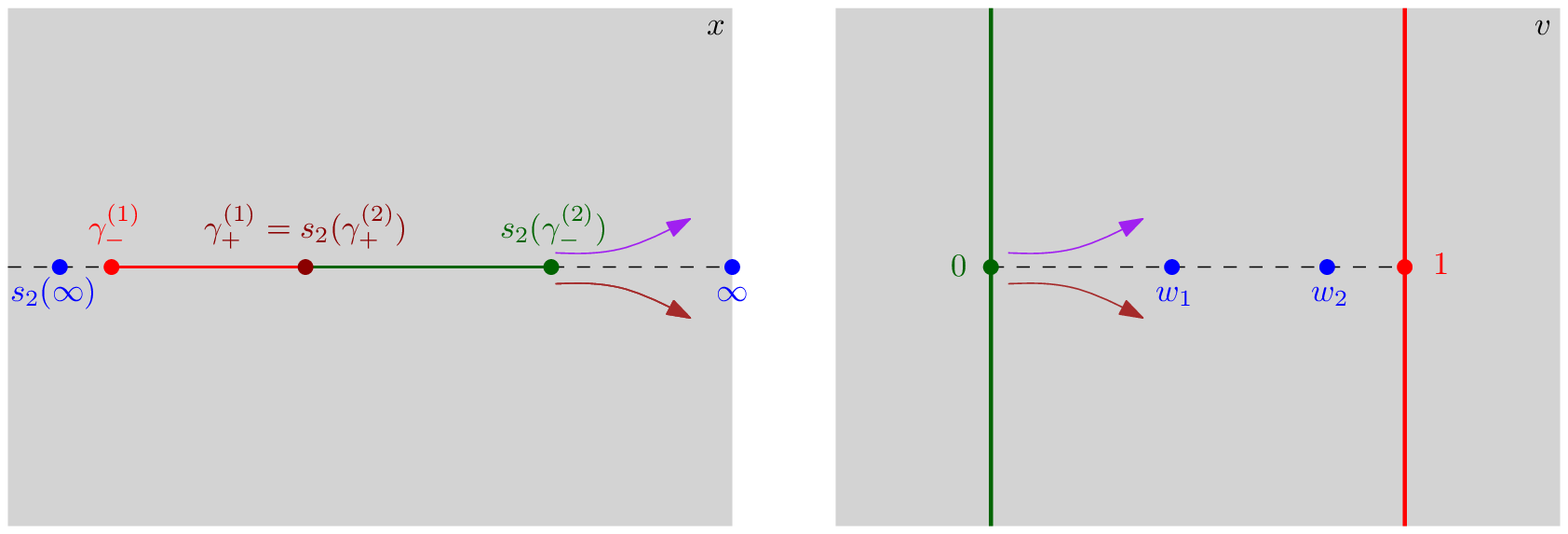}{13.cm}
\figlabel\trigoparam

\noindent When Eqn.~\overl\ is satisfied, the elementary function $\zeta_b(v)$ reduces to
\eqn\zetl{\zeta_b(v) = 2{\rm i}\pi\,{e^{{\rm i}\pi(b - 1)v} \over e^{{\rm i}\pi v} - e^{-{\rm i}\pi v}}.}
It is indeed easy to check that the right-hand side matches the required properties \polez.
Besides, the elliptic parametrization reduces to a trigonometric one, see Fig.~\trigoparam,
\eqn\aram{x(v) = \lambda - \delta\,{\cos\pi v - 1 \over \cos \pi v - \cos\pi w_1},}
where the three parameters $w_1$, $\lambda$ and $\delta$ are related to the position of the cuts via
\eqn\gam{\gamma^{(1)}_+ = s_2(\gamma^{(2)}_+)=\lambda - \delta,\qquad \gamma^{(1)}_- = \lambda-{2\delta \over 1 + \cos\pi w_1},\qquad
s_2(\gamma^{(2)}_-)=\lambda.}
The point $x = \gamma^{(1)}_+$ is now mapped to $v = -{\rm i}T = -{\rm i}\infty$, and we have
\eqn\gamB{x(v) \mathop{=}_{v \rightarrow -{\rm i}\infty}  \gamma^{(1)}_+ - 2\delta(\cos\pi w_1 - 1)e^{-{\rm i}\pi v}.}
The value of $w_2$ also fixed in terms of $(w_1,\lambda,\delta)$ since, from Eqn.~\polpot, we have $x(w_2) = s_2(\infty)$,
namely
\eqn\gam{ \lambda - \delta\,{\cos\pi w_2 - 1 \over \cos \pi w_2 - \cos\pi w_1}= {a \over (a^2 - 1)h^{(1)}}.}
We start with three useful observations:

\medskip
\item{$\bullet$} $w_1$ ranges over $]0,1[\cup\, {\rm i}{\bf R}_+^*$, and the condition that $\gamma^{(1)}_+>\gamma^{(1)}_-$ implies that $\delta(1 - \cos\pi w_1)$ remains positive.

\item{$\bullet$} We note that the asymptotic behaviors
\eqn\infifi{\eqalign{x'(v) & \mathop{\sim}_{v \rightarrow -{\rm i}\infty} 2{\rm i}\pi\,\delta (\cos\pi w_1 - 1)\,e^{-{\rm i}\pi v}, \cr x'(v + 1) & \mathop{\sim}_{v \rightarrow -{\rm i}\infty} -2{\rm i}\pi\,\delta (\cos\pi w_1 - 1)\,e^{-{\rm i}\pi v},}}
have opposite signs, and that $s_1$ is locally decreasing on the real line. Hence, the relative sign of $\rho^{(1)}(x(v + 1))$ and $\rho^{(2)}(s_1(x(v)))$ near $v \rightarrow -{\rm i}\infty$ is also the relative sign of $\rho^{(1)}(x(v + 1))x'(v + 1)$ and $\rho^{(2)}(s_1(x(v)))\,s_1'(x(v))\,x'(v)$.

\item{$\bullet$} For bookkeeping, we note the asymptotic expansions deduced from \zetl
\eqn\zetKL{\zeta_b(v) \mathop{=}_{v \rightarrow -{\rm i}\infty} \,\,2{\rm i}\pi\sum_{k \geq 0} e^{-{\rm i}\pi(2 - b + 2k)v},\qquad \zeta_{b}(-v) \mathop{=}_{v \rightarrow -{\rm i}\infty} \, - 2{\rm i}\pi\sum_{k \geq 0} e^{-{\rm i}\pi(b + 2k)v}.}

\medskip

Among the four conditions \POL\ and \POLQ, the two conditions that $\rho^{(1)}(x(v + 1))$ and $\rho^{(2)}(s_1(x(v)))$ vanish at $v = -{\rm i}\infty$ (second lines with $T\to \infty$) are better rephrased by saying that the coefficient in front of the terms of order $e^{-{\rm i}\pi bv}$ (arising from $\zeta_b(-v \pm w_j)$) and $e^{-{\rm i}\pi(1 - b)v}$ (arising from $\zeta_{\tilde{b}}(-v \pm w_j)$) must vanish, since they would lead to divergent terms in the densities (see \BBGa\ for a similar
argument). This being enforced, the dominant order in the densities is $e^{-{\rm i}\pi (1+b)v}$ and arises from terms involving $\zeta_{\tilde{b}}(v \pm w_j)$, but Eqn.~\densing\ and the observations we just made imply that this term comes with an opposite sign in $\rho^{(1)}$ and $\rho^{(2)}$. Thus, the positivity of the densities may only be achieved
if we impose an {\it extra relation}, namely that the coefficient in front of this term vanish. Then, the dominant order in the spectral densities when $v\to -{\rm i}\infty$ 
is $e^{-{\rm i}\pi (2-b)v}$ and comes from terms involving $\zeta_{b}(v \pm w_j)$, exactly as for a non-critical $O(n)$ model
(with $n=2\cos \pi b$) in its dense phase. The exponents of our non-generic critical twofold loop model are therefore generically the same as those of a usual loop model in its dense phase. Extra cancellations may drive these exponents to other values: 
if the coefficient in front of the dominant order $e^{-{\rm i}\pi (2-b)v}$ for one of the spectral densities happens to vanish at some values of the parameters
of the model, the dominant order for this spectral density becomes in practice $e^{-{\rm i}\pi (2+b)v}$ (arising from the sub-dominant order in $\zeta_b(-v \pm w_j)$),
leading to exponents whose values are now characteristic of the dilute phase of the $O(n)$ model.

To summarize, assuming a non-generic critical behavior imposes $\gamma^{(1)}_+ = s_2(\gamma^{(2)}_+)$, together with the ``extra relation'' that we just discussed. 
The non-generic critical variety is therefore of codimension $2$.

\subsec{Strategy for the fully-packed model}

From now on, we restrict ourselves to the fully-packed model. If the weight per loop $n$ and the bending energy $a$ are considered as fixed, we may parametrize 
the non-generic critical variety by expressing $h^{(1)}$ and $h^{(2)}$ on this variety as functions
of $u^{(1)}$ and $u^{(2)}$, or equivalently as functions of the reduced variables
\eqn\other{u = \sqrt{u^{(1)}u^{(2)}},\qquad r = {u^{(1)} \over u^{(2)}}.}
Carrying out this program is rather technical, so let us explain first our strategy. It is easily seen from Eqns.~\largea-\otherpole\ that,
in the fully-packed case, $\omega^{(i)}(v)$ has a triple pole at $w_1$ and $w_2$. On top of the residues known by Eqn.~\alphA\ (or Eqn.~\alphB\ for $a=1$), there are four unknown Laurent coefficients 
$(\alpha_{i|j})_{i,j = 1,2}$ (or $(\beta_{i|j})_{i,j = 1,2}$). Their expression in terms of $w_1$, $w_2$ and the parameters of the model can be derived directly from Eqn.~\otherpole\ (or Eqn.~\behlargxaun). We prefer to invert these relations to write $h^{(1)}$, $h^{(2)}$ and the position of the cuts in terms of $w_1,w_2$ and the Laurent coefficients. On the other hand, the $4$ equations ensuring that the densities vanish at the edges, can be viewed as a $4 \times 4$ linear system determining the $4$ Laurent coefficients in terms of $w_1,w_2,u,r$. Thus, we get expressions for $h^{(1)}, h^{(2)}$ and the position of the cuts in terms of $w_1,w_2,u,r$. Furthermore, using the solution of the $4 \times 4$ system to enforce the {\it extra relation}, we will obtain a first relation between $w_1,w_2$ and $r$. Enforcing Eqn.~\gam\ gives a second relation between $w_1,w_2$ and $r$ (for $a=1$, it is replaced by the even simpler requirement that $w_1=w_2$). Hence, $w_1$ and $w_2$ are entirely fixed by the value of $r$. Eventually, we check for the local positivity of the densities at $\gamma^{(i)}_{\pm}$, which may rule out certain values of parameters for the solution, and we check numerically for some admissible values that the densities are indeed globally positive.

In this way, we can arrive at a full analytical solution of the non-generic critical variety in the case $a = 1$, and a semi-analytical one in the case $a \neq 1$. Although $a = 1$ can be obtained by taking the limit $a \rightarrow 1$ (which is regular), we prefer to treat this case separately since it is less cumbersome than the general case.

\subsec{Fully-packed model without bending energy $(a = 1)$}

Recall first that $a = 1$ corresponds to $w_1 = w_2$. We use the set of functions introduced in Eqn.~\YdefA\ and the differential operators of Eqn.~\diffredef, and solve the $4 \times 4$ system determining the position of the cuts for the Laurent coefficients $(\beta_{i|j})_{i,j = 1,2}$. It can be written as
\eqn\sysfour{\eqalign{{\cal D}_1\big[\sin\pi b(1 -  w_1)\big] + {\cal D}_2\big[\sin\pi b w_1\big] & = 0, \cr
{\cal D}_1\big[\sin\pi(1 - b)(1 - w_1)\big] - {\cal D}_2\big[\sin\pi(1 - b)w_1\big] & = 0, \cr
{\cal D}_1\Big[{\cos\pi(1 - b)(1 - w_1) \over \sin\pi w_1}\Big] + {\cal D}_2\Big[{\cos\pi(1 - b)w_1 \over \sin\pi w_1}\Big] & = 0, \cr
{\cal D}_1\Big[{\cos\pi b(1 - w_1) \over \sin\pi w_1}\Big] - {\cal D}_2\Big[{\cos\pi b w_1 \over \sin\pi w_1}\Big] & = 0.}}
Then, the {\it extra relation} reads
\eqn\fifths{{\cal D}_1\big[\sin\pi(1 + b)(1 - w_1)\big] - {\cal D}_2\big[\sin\pi(1 + b)w_1\big] = 0,}
and if we plug in the solution of Eqn.~\sysfour, we arrive to the surprisingly simple relation
\eqn\ratioB{r = {\kappa_{b}(w_1) \over \kappa_b(1 - w_1)},\qquad\kappa_b(w) = b\,\cos\pi b w\,\sin\pi w - \sin\pi bw\,\cos\pi w.}
which fixes $w_1$ as a function of $r$ only.
It is easy to see that, when $r$ assumes positive values and $b \in ]0,1/2[$, Eqn.~\ratioB\ has a unique solution $w^*_1(r) \in ]0,1[$, and no solution in ${\rm i}{\bf R}_+$. Then, we find that the spectral densities behaves near $\gamma^{(i)}_{+}$ as
\eqn\denspo{\eqalign{\rho^{(1)}(x(v + 1)) & \mathop{\sim}_{v \rightarrow -{\rm i}\infty} {1 \over 2\pi}\Big({4u\,\sin\pi b \over b\,g_b(w_1)}\Big)^{1/2}\,\Big({\sin\pi w_1 \over \sin\pi b(1 - w_1)}\Big)^{1/2}\,e^{-{\rm i}\pi(1 - b)v}, \cr
\rho^{(2)}(s_1(x(v))) & \mathop{\sim}_{v \rightarrow -{\rm i}\infty} {1 \over 2 \pi}\Big({4u\,\sin\pi b \over b\,g_b(w_1)}\Big)^{1/2}\,\Big({\sin\pi w_1 \over \sin\pi bw_1}\Big)^{1/2}\,e^{-{\rm i}\pi(1 - b)v},}}
and near $\gamma^{(i)}_{-}$ as
\eqn\denspoB{\eqalign{\rho^{(1)}(x) & \mathop{\sim}_{x \rightarrow \gamma^{(1)}_-} {b \over \sqrt{2}\, \pi}\Big({4u\,\sin\pi b \over b\,g_b(w_1)}\Big)^{1/4}\,{(x - \gamma_-^{(1)})^{1/2} \over (\sin\pi w_1)^{1/4}(\sin\pi b(1 - w_1))^{3/4}}, \cr
\rho^{(2)}(x) & \mathop{\sim}_{x \rightarrow \gamma^{(2)}_-} {b \over \sqrt{2}\, \pi}\Big({4u\,\sin\pi b \over b\,g_b(w_1)}\Big)^{1/4}\,{(x - \gamma_-^{(2)})^{1/2} \over (\sin\pi w_1)^{1/4}(\sin\pi bw_1)^{3/4}}.}}
Hence the spectral densities are locally positive at the edges of the cuts for all values of $w_1 \in ]0,1[$, and we checked numerically for several values of $b$ and $w_1$ that they remain globally positive along the cuts.

\fig{Non-generic critical values of $h^{(1)}$ and $h^{(2)}$ as functions of $r =u^{(1)}/u^{(2)}$, for $b = 0.4, a = 1, u = 1$.}{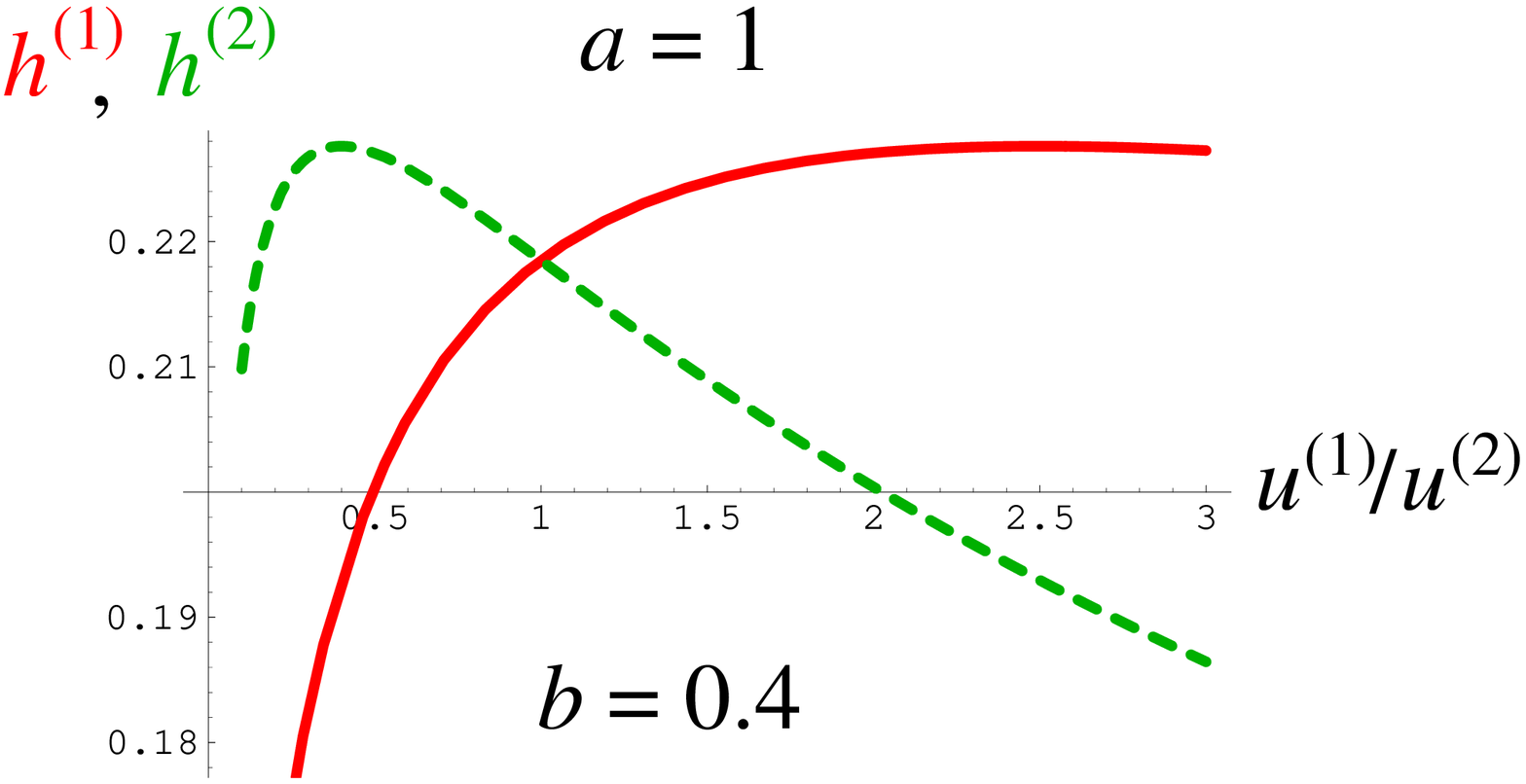}{8.cm}
\figlabel\honetwo

Next, as explained in Section~5.2, using Eqns.~\behlargxaun-\omegadef-\alphB\ and the solution of Eqns.~\sysfour-\fifths, we arrive after some algebra at
\eqn\oye{\eqalign{h^{(1)} & = \Big({4u\,\sin\pi b \over b\,g_b(w_1)}\Big)^{-1/2}\,{1 \over b\,\sin\pi b}\,{\sqrt{\sin\pi b(1 - w_1)}\sin\pi b w_1 \over \sqrt{\sin\pi w_1}}, \cr
h^{(2)} & =  \Big({4u\,\sin\pi b \over b\,g_b(w_1)}\Big)^{-1/2}\,{1 \over b\,\sin\pi b}\,{\sqrt{\sin\pi b w_1}\sin\pi b(1 - w_1) \over \sqrt{\sin\pi w_1}}, \cr
\gamma^{(1)}_+ & =  \Big({4u\,\sin\pi b \over b\,g_b(w_1)}\Big)^{1/2}\,{b\,\cos\pi b(1 - w_1)\sqrt{\sin\pi w_1} \over \sqrt{\sin\pi b(1 - w_1)}},\cr
\gamma^{(1)}_- & = \Big({4u\,\sin\pi b \over b\,g_b(w_1)}\Big)^{1/2}\,{b\,\sin\pi w_1\cos\pi b(1 - w_1) - (1 - \cos\pi w_1)\sin\pi b(1 - w_1) \over \sqrt{\sin\pi w_1\sin\pi b(1 - w_1)}}, \cr
\gamma^{(2)}_+ & = \Big({4u\,\sin\pi b \over b\,g_b(w_1)}\Big)^{1/2}\,{b\,\cos\pi bw_1\sqrt{\sin\pi w_1} \over \sqrt{\sin\pi bw_1}}, \cr
\gamma^{(2)}_- & = \Big({4u\,\sin\pi b \over b\,g_b(w_1)}\Big)^{1/2}\,{b\,\cos\pi bw_1\sin\pi w_1 - (1 + \cos\pi w_1)\sin\pi bw_1 \over \sqrt{\sin\pi w_1\sin\pi bw_1}},}}
where we have set
\eqn\gbdef{g_b(w) = \sqrt{\kappa_b(w_1)\kappa_b(1 - w_1)}.}
This is a parametrization of the non-generic critical variety in terms of $u$, and $w_1$ which is itself uniquely determined by $r$ (Eqn.~\ratioB).
Notice that the ratio $h^{(1)}/h^{(2)}$ has a simple expression in terms of $w_1$, namely
\eqn\atio{{h^{(1)} \over h^{(2)}} = \sqrt{\sin\pi b w_1 \over \sin\pi b(1 - w_1)}.}
For illustration, we have plotted in Fig.~\honetwo\ the values of $h^{(1)}$ and $h^{(2)}$, as given by \ratioB, \gbdef\ and \oye, 
at $b = 0.4$, $u=1$ and for varying $r$, i.e. $u^{(1)}=\sqrt{r}=1/u^{(2)}$.

\medskip

\noindent $\bullet$ In the limit case $n \rightarrow 0$ (i.e. $b = 1/2$), we obtain the simple relation
\eqn\ovenun{r = {\rm tan}^3(\pi w_1/2),} 
so that we may write, in terms of the parameters $u,r$,
\eqn\ovenzero{h^{(1)} = {1 \over 2}\,u^{-1/2}\,{r^{-1/4} \over 1 + r^{-2/3}},\qquad  h^{(2)} = {1 \over 2}\,u^{-1/2}\,{r^{1/4} \over 1 + r^{2/3}},}
and the cuts are symmetric with respect to $0$:
\eqn\cutzero{\gamma^{(1)}_{\pm} = \pm 2\,u^{1/2}\,r^{1/4},\qquad \gamma^{(2)}_{\pm} = \pm 2\,u^{1/2}\,r^{-1/4}.}

\medskip

\noindent $\bullet$ The symmetric case (i.e. $r = 1$) is particularly simple as Eqn.~\ratioB\ yields $w_1=1/2$ so that, from \oye\ or \atio,
\eqn\symh{h^{(1)}=h^{(2)}={1\over 2\sqrt{2 u}\sqrt{2+n}},}
as already known from \KosGau. The model is then fully red-green-symmetric.

\medskip

\fig{Critical couplings as a function of $Q$ for the Potts model (plain line), and for the symmetric model (which correspond to self-dual values, dashed line). Here, $u = 1$ is assumed.}{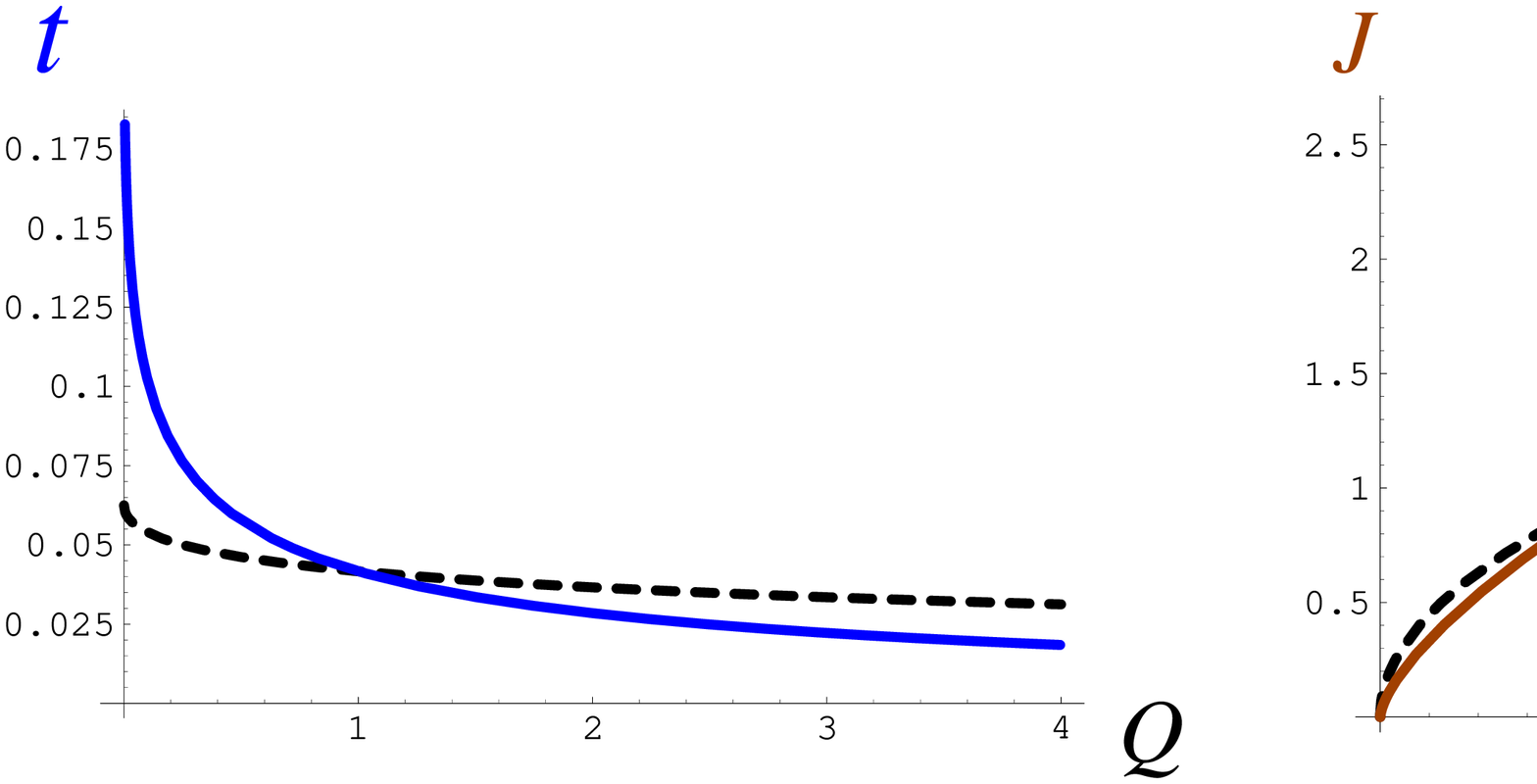}{13.cm}
\figlabel\tJpotts

\noindent $\bullet$ Let us now return to the Potts model with $\mu_b=\mu_f=1$, i.e. $r = 2\cos\pi b$ and $u = \sqrt{2\cos\pi b}$. We get 
from \ratioB\ and \oye\ expressions for the critical couplings $t$ and $J$ via
\eqn\jtjt{t=(h^{(2)})^2,\qquad J=\sqrt{Q} (h^{(1)}/h^{(2)})^2,}
as obtained by inverting Eqn.~\triangleweights. We have plotted these values as functions of $Q = 4\cos^2\pi b$ in Fig.~\tJpotts. We note that, for $Q \neq 1$, those are {\it not} the self-dual values, which correspond instead to a symmetric model with $u^{(1)}=u^{(2)}=1$, hence $h^{(1)} = h^{(2)}$. These self-dual values for $J$ and $t$ are plotted for comparison in Fig.~\tJpotts\ in dashed lines. We insist here on the fact that the Potts model on general random maps is not self-dual for $Q\neq 1$ (and the value of its critical couplings cannot be obtained by a simple duality argument), although its critical point clearly lies in the same universality class as the domain-symmetric
model. This is in contrast with the Potts model on regular lattices, where the self-duality argument is commonly used to determine critical values of the couplings.

\subsec{Fully-packed model with bending energy $(a \neq 1)$}

When $a \neq 1$, we use the Laurent coefficients $(\alpha_{i|j})_{i,j = 1,2}$. By comparing Eqns.~\alphA, \aram, \otherpole\ and \omegadef, we first obtain explicit expressions for $h^{(1)}$, 
$h^{(2)}$, $\lambda$ and $ \delta$, and thus for the position of the cuts, as  functions of $w_1$, $w_2$ and $\alpha_{i|j}$. In particular we get:
\eqn\inverA{\eqalign{h^{(1)} & = {1 \over 2\sqrt{4 - n^2}}\,{1 \over a(1 - a^2)}\,{\cos\pi w_1 - \cos\pi w_2 \over \sin\pi w_1\sin^2\pi w_2} \cr
& \times {\pi^2 (\alpha_{2|2}/2)(\cos^2\pi w_2 + \cos\pi w_2\cos\pi w_1 - 2) + \pi\alpha_{2|1}\sin\pi w_2(\cos\pi w_1 - \cos\pi w_2) \over \pi^2(\alpha_{2|2}/2)\,\sqrt{\pi^2\alpha_{1|2}/2}}, \cr
h^{(2)} & = {2 \over \sqrt{4 - n^2}}\,{a \over 1 - a^2}\,\sin\pi w_2(\cos\pi w_1 - \cos\pi w_2) \cr
& \times {\sqrt{\pi^2 \alpha_{2|2}/2} \over \pi^2(\alpha_{2|2}/2)(\cos^2\pi w_2 + \cos\pi w_2\cos\pi w_1 - 2) + \pi\alpha_{2|1}\sin\pi w_2(\cos\pi w_1 - \cos\pi w_2)}, \cr
\gamma^{(1)}_+ & = {\sqrt{4 - n^2} \over 2}\,{- \pi^2(\alpha_{1|2}/2){\rm cotan}(\pi w_1) - \pi\alpha_{1|1} \over \sqrt{\pi^2\alpha_{1|2}/2}}, \cr
\gamma^{(1)}_- & = {\sqrt{4 - n^2} \over 2}\,{\pi^2(\alpha_{1|2}/2)(2 - \cos\pi w_1) + \pi\alpha_{1|1}\sin\pi w_1 \over \sin\pi w_1\sqrt{\pi^2\alpha_{1|2}/2}}.}}
Since $s_1(x(v))$ is also a rational function of $\cos\pi v$ with a pole at $v = w_2$, it is possible to derive from Eqns.~\largea-\otherpole\ expressions for $h^{(1)},h^{(2)}$, together with the coefficients $({\tilde \lambda},{\tilde \delta})$ defined through
\eqn\sAxv{s_1(x(v)) = {\tilde \lambda} - {\tilde \delta}\,{\cos\pi v - 1 \over \cos\pi v - \cos\pi w_2},}
in terms of $w_1$, $w_2$ and $\alpha_{i|j}$. In this way, we find the alternative expressions
\eqn\inverB{\eqalign{h^{(1)} & = {2 \over \sqrt{4 - n^2}}\,{a \over (1 - a^2)}\,(\cos\pi w_1 - \cos\pi w_2)\sin\pi w_1 \cr
& \times {\sqrt{\pi^2\alpha_{1|2}/2} \over -\pi^2(\alpha_{1|2}/2)(\cos^2\pi w_1 + \cos\pi w_1\cos\pi w_2 - 2) + \pi\alpha_{1|1}\sin\pi w_1(\cos\pi w_1 - \cos\pi w_2)}, \cr
h^{(2)} & = {1 \over 2\sqrt{4 - n^2}}\,{1 \over a(1 - a^2)}\,{\cos\pi w_1 - \cos\pi w_2 \over \sin^2\pi w_1\sin\pi w_2} \cr
& \times {-\pi^2(\alpha_{1|2}/2)(\cos^2\pi w_1 + \cos\pi w_1\cos\pi w_2 - 2) + \pi\alpha_{1|1}\sin\pi w_1(\cos\pi w_1 - \cos\pi w_2) \over \pi^2(\alpha_{1|2}/2)\sqrt{\pi^2\alpha_{2|2}/2}}, \cr
\gamma^{(2)}_+ & = {\sqrt{4 - n^2} \over 2}\,{\pi^2(\alpha_{2|2}/2){\rm cotan}(\pi w_2) + \pi \alpha_{2|1} \over \sqrt{\pi^2\alpha_{2|2}/2}}, \cr
\gamma^{(2)}_- & = {\sqrt{4 - n^2} \over 2}\,{\pi^2(\alpha_{2|2}/2)(2 + \cos\pi w_2) - \pi\alpha_{2|1}\sin\pi w_2 \over \sin\pi w_2\sqrt{\pi^2(\alpha_{2|2}/2)}}.}}
Let us stress that imposing Eqn.~\gam\ is then fully equivalent to demanding that the two above expressions \inverA\ and \inverB\ for $h^{(1)}$ (or for $h^{(2)}$) are 
indeed equal.

\fig{The function $w_2 = \phi_{b;r}(w_1)$ for $b = 0.3$ and several values of $r$. For $r = 1$, we have $\phi_{b;r = 1}(w_1) = 1 - w_1$ (red line). Above this line, we have plotted $\phi_{b;r}$ for $r = 3/2, 2, 3, 5, 7, 10$ and $20$. Under this line, we find $\phi_{b;r}$ for the inverse values ($2/3, 1/2$, etc.) by axial symmetry. The lower right quadrant corresponds to $w_1 \in ]0,1[$ and $w_2 \in ]0,1[$. In the lower left quadrant, $w_1$ becomes pure imaginary, more precisely in ${\rm i}{\bf R}_+^*$, while $w_2$ remains in $]0,1[$. In the upper right quadrant, $w_1$ remains in $]0,1[$ while $(1 - w_2)$ belongs to ${\rm i}{\bf R}_+^*$. Finally, in the upper right quadrant, $w_1$ and $(1 - w_2)$ both belong to  ${\rm i}{\bf R}_+^*$. Notice that $\phi_{b;r}$ is continuous across the axes $w_1 = 0$ and $(1 - w_2) = 0$. Once $r$ is fixed, hence $(w_1,w_2)$ lies on one of the lines in the graphic, the value of $a$ selects one particular point on this line according to \gam. For instance, $a = 1$ corresponds to points at the intersection of the fixed $r$-lines and the dashed magenta line $w_1 = w_2$, increasing $a$ corresponds to intersection points moving towards the lower right corner. Furthermore, the positivity conditions exclude a domain (here plotted qualitatively in green) in this lower right quadrant, without affecting other the quadrants. This leads to the existence of some maximal value $(w_1)_{\max}[r]$ admissible on the non-generic critical line, which in turn fixes a maximal value $a_{\max}[r]$. Changing the value of $b$ does not modify qualitatively (and even quantitatively when $b$ is between $0.1$ and $0.4$) the graphs above, except for $b \rightarrow 1/2$ (i.e. $n \rightarrow 0$) where the forbidden green domain is pushed towards the axes. In other words, the positivity condition is satisfied everywhere at $b = 1/2$.}{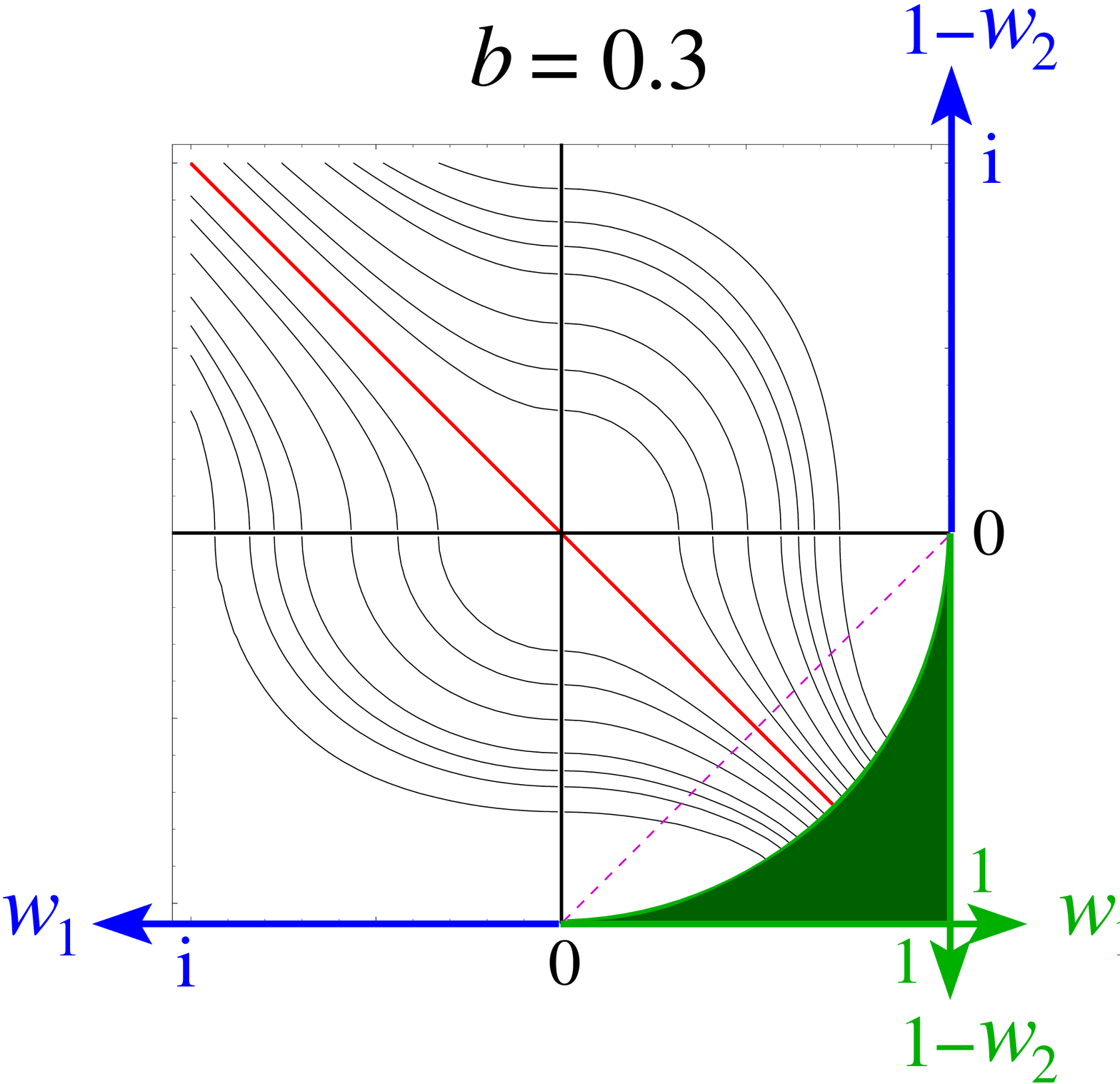}{9.5cm}
\figlabel\suau

Now, we can use the set of functions introduced in Eqn.~\Zde\ together with the differential operators of Eqn.~\diffop, and obtain the Laurent coefficients $\alpha_{i|j}$
by solving the $4 \times 4$ system ensuring that the densities vanish at the edges. The system itself reads
\eqn\condAA{\eqalign{{\cal D}_1\big[\cos\pi b w_1\big] + {\cal D}_2\big[\cos\pi b (1 - w_2)\big] & = 0, \cr
{\cal D}_1\big[\cos\pi(1 - b)w_1\big] - {\cal D}_2\big[\cos\pi(1 - b)(1 - w_2)\big] & = 0, \cr
{\cal D}_1\Big[{\sin \pi bw_1 \over \sin\pi w_1}\Big] - {\cal D}_2\Big[{\sin \pi b(1 - w_2) \over \sin\pi(1 - w_2)}\Big] & = 0, \cr {\cal D}_1\Big[{\sin \pi(1 - b)w_1 \over \sin\pi w_1}\Big] + {\cal D}_2\Big[{\sin\pi(1 - b)(1 - w_2) \over \sin\pi(1 - w_2)}\Big] & = 0,}}
and its solution yields cumbersome expressions for $\alpha_{i|j}$ in terms of $w_1,w_2,u,r$. We may plug these expressions back in the {\it extra relation}, which can be written here as
\eqn\condBB{{\cal D}_1\big[\cos\pi(1 + b)w_1\big] - {\cal D}_2\big[\cos\pi(1 + b)(1 - w_2)\big] = 0.}
The result is that Eqn.~\condBB\ becomes equivalent to a relation of the form
\eqn\ratio{r = {\kappa_b(w_1,1 - w_2) \over \kappa_b(1- w_2,w_1)},}
where $\kappa_b(w,w')$ is a complicated trigonometric function which we will not copy here. We find graphically that, for a given $r > 0$ and $w_1 \in ]0,1[\,\cup\,{\rm i}{\bf R}_+^*$, only two solutions for $w_2$ exist in $]0,1[\,\cup\,(1 - {\rm i}{\bf R}_+^*)$. More precisely, there exists two continuous functions $(\phi_{b;r}^{(k)})_{k = 1,2}$ such that $(w_1,\phi_{b;r}^{(k)}(w_1))$ is a solution of Eqn.~\ratio. However, we find that only one of them can yield a spectral density $\rho^{(1)}$ which is locally positive near $\gamma^{(1)}_+$. In this case, local positivity implies that $w_1 < (w_1)_{\max}[r]$ or $w_1 \in {\rm i}{\bf R}_+^*$. We call this function $\phi_{b;r}$. Numerically, it does not depend very much on $b$ far from $0$ or $1/2$. We give its plot for $b = 0.3$ and several values of $r$ in Fig.~\suau. If $r = 1$, $\phi_{b;1}(w_1) = 1 - w_1$ is an obvious solution, which indeed satisfies the local positivity constraint, provided $w_1 < (w_1)_{\max}[r = 1]$ or $w_1 \in {\rm i}{\bf R}_+^*$. We shall return to this
case in Section 5.5 below.

\fig{Critical values of $h^{(1)}$ and $h^{(2)}$ as a function of $r=u^{(1)}/u^{(2)}$ for several values of $a$, and the values $b = 0.4,u = 1$. In (b), we have chosen the value $a = 1.7$ around which an exchange between $h^{(1)}$ and $h^{(2)}$ occurs (compare Fig.~\honetwo\ and (a) here).}{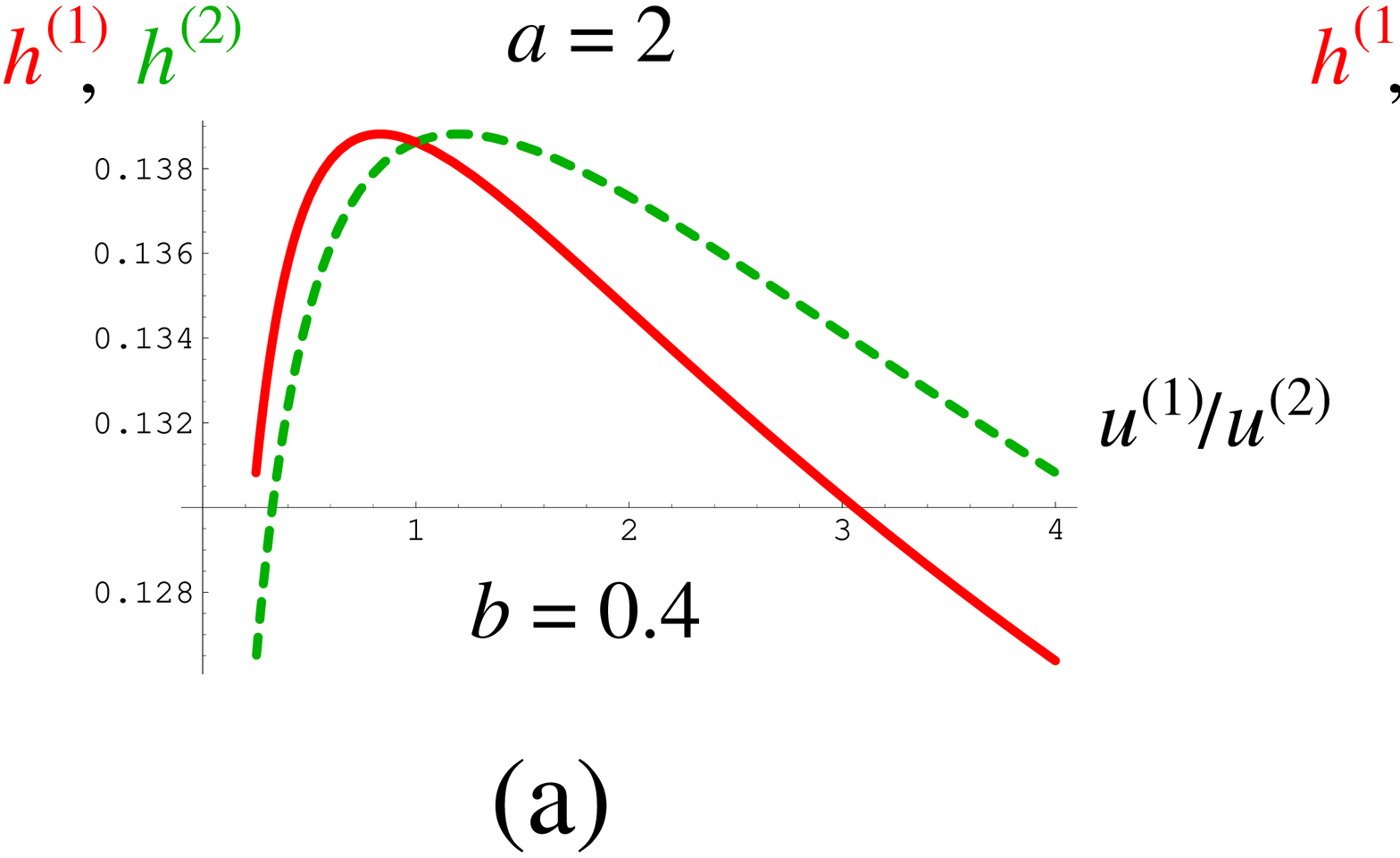}{14.cm}
\figlabel\honetwoatwo

To summarize, Eqn.~\ratio\ and Eqn.~\gam\ (or equivalently the matching of the two expressions for $h^{(1)}$ in \inverA\ and \inverB) provide two equations determining entirely $w_1$ and $w_2$ as a function of $r$ and $a$. For illustration, we have plotted in Fig.~\honetwoatwo-(a) the values of $h^{(1)}$ and $h^{(2)}$
at $b=0.4$, $u=1$ and for varying $r$ (i.e. $u^{(1)}=\sqrt{r}=1/u^{(2)}$) for the particular value $a=2$. Surprinsingly,
we see that in this case $h^{(2)} > h^{(1)}$ whenever $r>1$, in contrast with what we had for $a=1$.
There exists therefore an intermediate value of $a$ where this ``exchange'' takes place. For $b=0.4$,
we find that this exchange occurs for $a\sim 1.7$, as displayed in Fig.~\honetwoatwo-(b).

\medskip

\fig{Critical values of $h^{(1)}$ and $h^{(2)}$ as a function of $r=u^{(1)}/u^{(2)}$ for several values of $a$, in the limit $n \rightarrow 0$, and for $u = 1$. The exchange between $h^{(1)}$ and $h^{(2)}$ when passing from (a) to (b) occurs at $a = 4/3$ (plotted in (c)).}{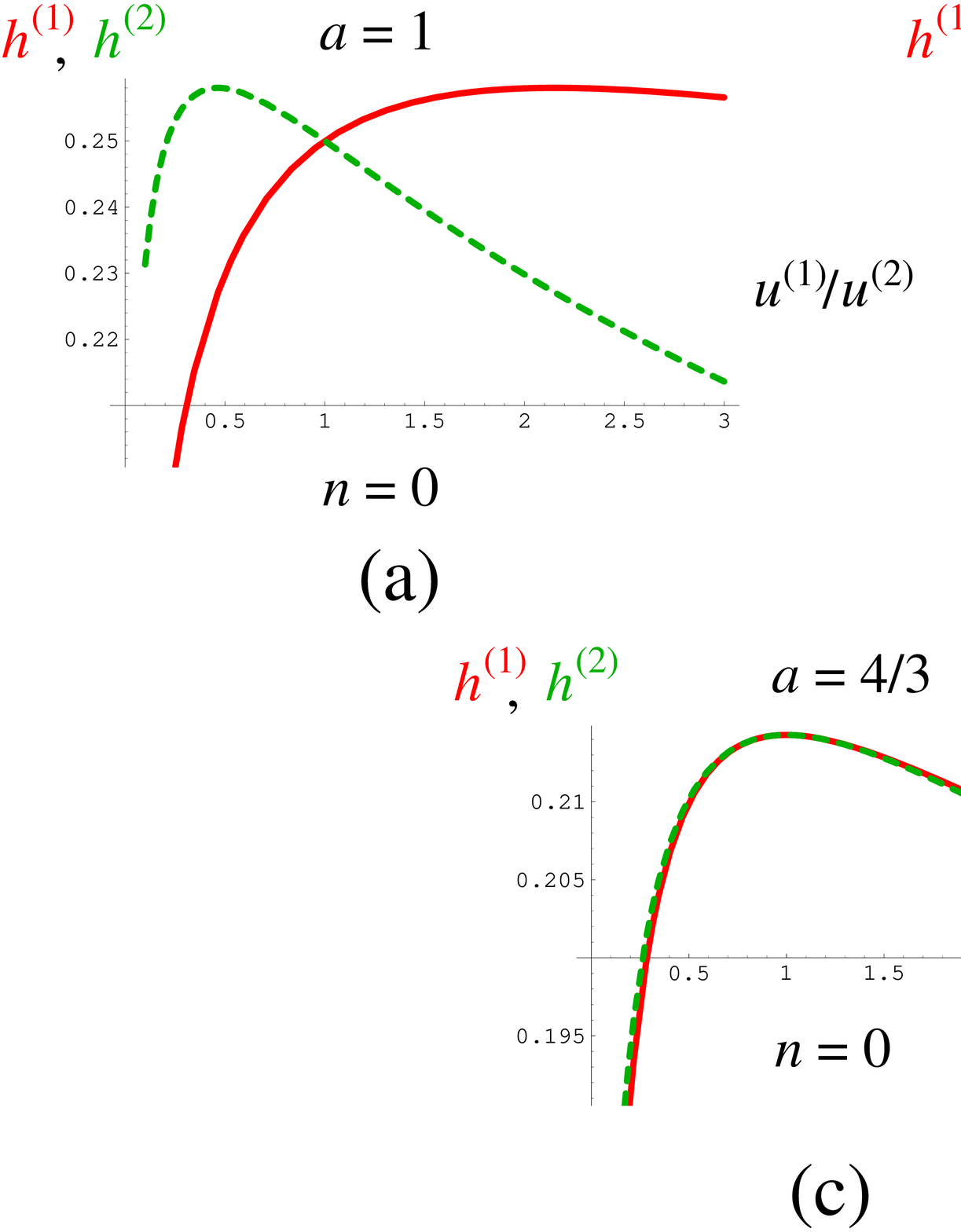}{14.cm}
\figlabel\honetwoatwozero

\noindent $\bullet$ The results simplify drastically in the limit $n \rightarrow 0$. We find in this case the two conditions
\eqn\uun{r = {\sin^3(\pi w_2/2) \over \cos^3(\pi w_1/2)},\qquad 4a^2(1 + \cos\pi w_1)(1 - \cos\pi w_2) = 4 - (\cos\pi w_1 + \cos\pi w_2)^2,}
which fix $w_1$ and $w_2$ in terms of $r$ and $a$. More explicitly, we find
\eqn\wde{1 + \cos(\pi w_1) = {2 \over \eta_{a,r}},\qquad 1 - \cos(\pi w_2) = {2\,r^{2/3} \over \eta_{a,r}},}
where
\eqn\etadef{\eta_{a,r} = \sqrt{4a^2r^{2/3} + (1 - r^{2/3})^2}.}
This leads eventually to the following parametrization of the critical variety
\eqn\uub{\eqalign{h^{(1)} & = {u^{-1/2}\,r^{-7/12} \over 4(a^2 - 1)}\,(\eta_{a,r} - 1 - r^{2/3})\Big({\eta_{a,r} - 1 + r^{2/3} \over \eta_{a,r} + 1 - r^{2/3}}\Big)^{1/2},\cr
h^{(2)} & = {u^{-1/2}\,r^{-1/12} \over 4(a^2 - 1)}\,(\eta_{a,r} - 1 - r^{2/3})\,\Big({\eta_{a,r} + 1 - r^{2/3} \over \eta_{a,r} - 1 + r^{2/3}}\Big)^{1/2}, \cr
\gamma^{(1)}_{\pm} & = \pm 2u^{1/2}\,r^{1/4},\cr
\gamma^{(2)}_{\pm} & = \pm 2u^{1/2}\,r^{-1/4}.}}
Here, it turns out that $(w_1)_{\max}[r] = 1$, i.e. one can check explicitly that the spectral densities are locally positive at their edges for all positive values of $a,r$
(in there words, the forbidden green domain in Fig.~\suau\ is pushed toward the axes when $n\to 0$). For illustration, we have plotted in Fig.~\honetwoatwozero\ the values of $h^{(1)}$ and $h^{(2)}$ at $n=0$, $u=1$ and for varying $r$ (i.e. $u^{(1)}=\sqrt{r}=1/u^{(2)}$) for the particular values $a=1$, $a=2$ and $a=4/3$ at which an exchange between $h^{(1)}$ and $h^{(2)}$ occurs. Notice that, in the limit $a \rightarrow 1$, we recover the results of Eqn.~\ovenzero. Notice also that $\gamma^{(i)} = \pm 2\sqrt{u^{(i)}}$ is totally expected as $n \rightarrow 0$ corresponds to having a single loop separating a red and a green domain, each reduced to a tree, with respective weight per vertex $u^{(1)}$ and $u^{(2)}$. 

\medskip

\noindent $\bullet$ We point out that, at $n = 0$ strictly, we have $b = \tilde{b} = 1/2$, and therefore the dominant contribution to the spectral density comes from the sum of the coefficients in front of $\zeta_{\tilde{b}}(v \pm w_j)$ and $\zeta_b(v \pm w_j)$. So, we need this sum to be positive, rather than cancellation of the coefficient of $\zeta_{\tilde{b}}(v \pm w_j)$ alone. In other words, the {\it extra relation} introduced in Section~5.2 is no longer necessary, and the critical variety has codimension $1$. Its equation is obtained by demanding only $s_1(\gamma^{(1)}_+) = \gamma^{(2)}_+$, namely
\eqn\critit{1 - 2a\sqrt{u^{(1)}}\,h^{(1)}- 2a\sqrt{u^{(2)}}\,h^{(2)}\ + 4(a^2 - 1)\sqrt{u^{(1)}u^{(2)}}\,h^{(1)}h^{(2)} = 0,}
and the positivity condition is satisfied for all positive values of the parameters. Eqn.~\uub\ is simply the point on this line which satisfies the (no longer required) {\it extra relation}. From a combinatorial point of view, we do know if this point has a special meaning.

\subsec{The symmetric fully-packed model}

The symmetric fully-packed model corresponds to having $u^{(1)}=u^{(2)}=u$, i.e. $r=1$, and $h^{(1)}=h^{(2)}=h$. In this case, the condition \ratio\ yields $w_2=1-w_1$, and the problem reduces to the $O(n)$ model with bending energy studied in \BBGa, with moreover $g=0$ in the notations of that paper. We may therefore refer to the simpler analysis of \BBGa\ to deduce the value of $w_1$ (hence of $w_2$). From the expressions of \BBGa, we find that it is fixed by the condition
\eqn\atow{a={b\sin(2\pi w_1)(2 + \cos 2\pi bw_1) - \sin(2\pi bw_1)(2 + \cos 2\pi w_1) \over \sin ^2\pi w_1\,(b \sin 2\pi w_1-\sin 2\pi bw_1)}.}
Note that $w_1 \in [0,1]$ for $a\geq {2\over 5}(1+b^2)$ (with in particular $w_1=1/2$ for $a=1$), while $w_1$ is purely imaginary otherwise.
\fig{Critical line in the $(h,a)$ plane for the symmetric fully-packed model, at $b = 0.4$ and $u = 1$. In the blue part, $w_1$ is pure imaginary, while in the green part, it belongs to $]0,1$[.}{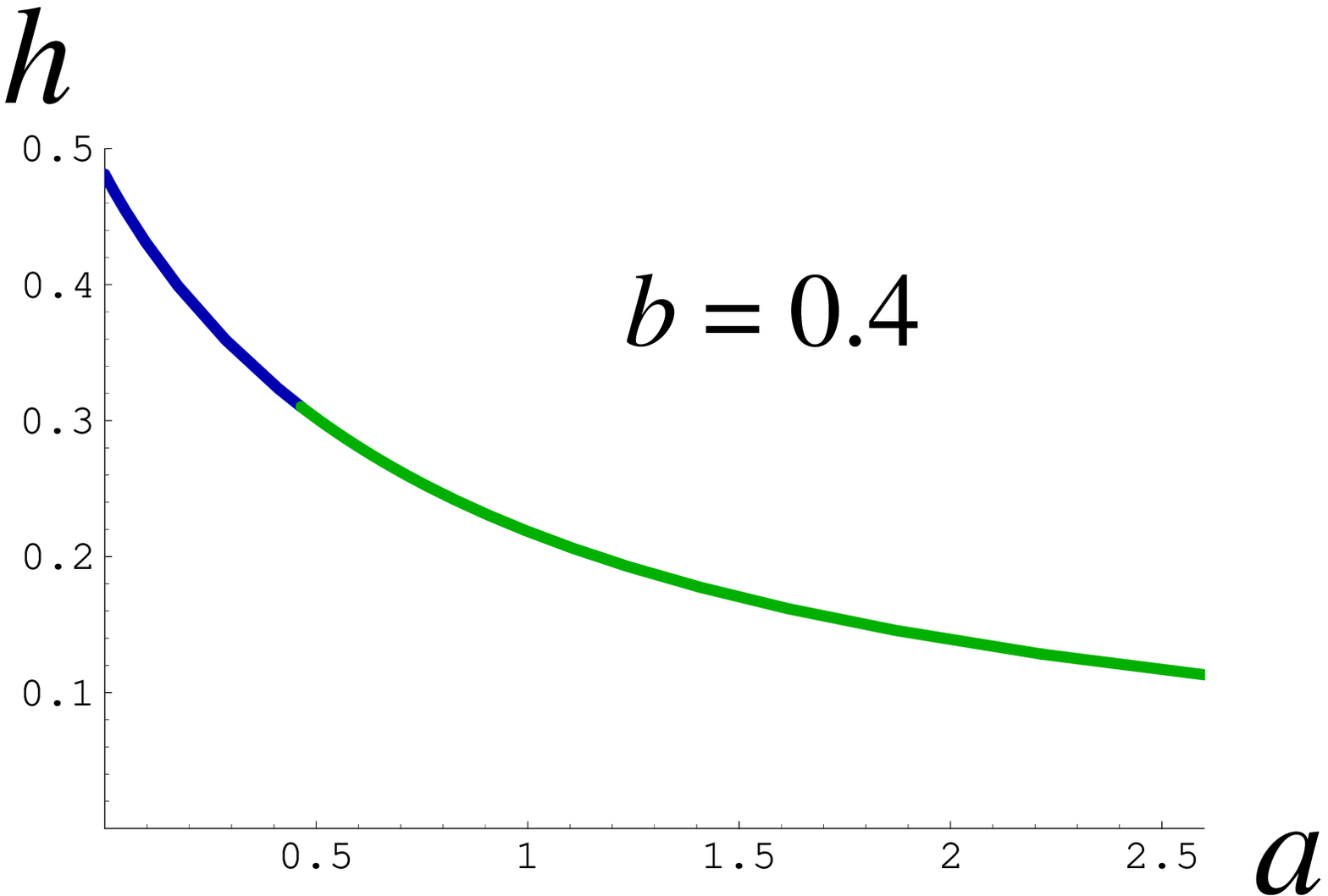}{8.cm}
\figlabel\hsymvsa

\medskip

The non-generic critical point is then found at the symmetric value
\eqn\valhsym{h^{(1)}=h^{(2)}=h= {\sqrt{H}\over \sqrt{2-n}|a^2-1| \sqrt{u}},}
with
\eqn\valH{H= {4b\cos ^2\pi  w_1\left(-b^2\sin^2\pi w_1\cos\pi w_1 - b\sin\pi w_1\sin 2\pi bw_1 + 2\cos\pi w_1\sin^2\pi bw_1\right) \over  \sin^3\pi w_1(b\sin 2\pi w_1 -\sin 2\pi bw_1)}.}
For illustration, we have plotted the value of $h$ versus $a$ for $b=0.4$ in Fig.~\hsymvsa. 
\fig{Maximal admissible value of $a$ for non-generic critical points in the symmetric fully-packed model, as a function of $n$. $a_{\max}$ ranges between $+\infty$ (at $n = 0$) to $2$ (at $n = 2$), and assumes the value $4$ at $n = 1$.}{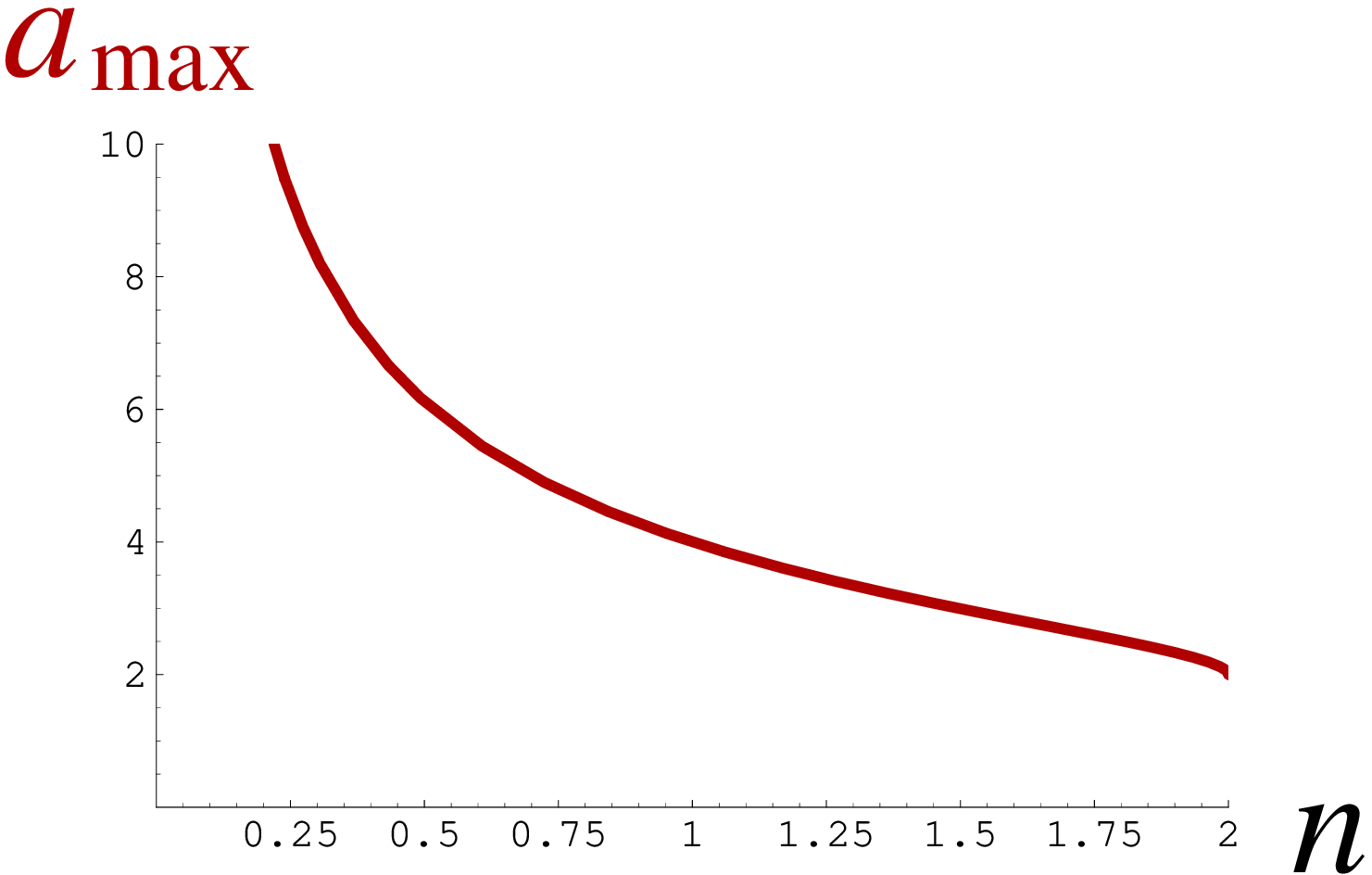}{8.cm}
\figlabel\amax
\noindent The condition that the spectral density $\rho^{(1)}$ be non-negative for $x\to (\gamma^{(1)}_+)^-$ reads $w_1\leq (w_1)_{\rm max} = (w_1)_{\max}[1]$ where $(w_1)_{\rm max} > 1/2$ is fixed by the explicit condition $f_b[(w_1)_{\max}] = 0$, with
\eqn\wonemax{\eqalign{f_b(w) & = b^2[-\cos\pi(6 - b)w - \cos\pi(4 - b)w + 2\cos\pi(2 - b)w + 2\cos\pi bw \cr 
& - \cos\pi(2 + b)w] + b[\cos\pi(6 - b)w + \cos\pi(4 - b)w - 4\cos\pi(2 - b)w - 4\cos\pi w \cr 
& + 3\cos\pi(2 + b)w + \cos\pi(b + 4)w] + 8\cos\pi w\sin\pi bw\sin\pi(3 - 2b)w.}}
From \atow, this condition amounts to the condition $a\leq a_{\rm max}$ with $a_{\rm max}$ given by \atow\ with 
$w_1=(w_1)_{\rm max}$. We have plotted in Fig.~\amax\ the value of $a_{\rm max}$ as a function of $n$. It 
decreases from $a_{\rm max}=\infty$ for $n=0$ ($b=1/2$) to $a_{\rm max}=2$ for $n=2$ (i.e. $b=0$).

\medskip
Note that, technically, when $a < a_{\rm max}$, the dominant order of the spectral densities when $v \rightarrow -{\rm i}\infty$ is given by the terms involving $\zeta_b(v \pm w_j)$. Since we obtain locally negative densities when $a > a_{\max}$, the coefficient of this (generically) dominant order should vanish at $a = a_{\max}$, yielding new exponents different from those of the dense phase. For the symmetric fully-packed model at $a = a_{\max}$, we expect that the new dominant order is given by the first subleading terms involving $\zeta_{b}(v \pm w_j)$, recovering in this way exponents of the dilute phase of the $O(n)$ model.

\medskip

\noindent $\bullet$ For $n=0$ ($b=1/2$), Eqns.~\atow-\valH\ reduce to
\eqn\atownzero{a={1\over 1+\cos \pi w_1},\quad H= {1\over 2}\left({\cos \pi w_1\over 1+ \cos \pi w_1}\right)^2,}
in agreement with \wde-\uub\ for $r=1$. We have then $(w_1)_{\rm max}=1$ and $a_{\rm max}=\infty$.

\medskip

\noindent $\bullet$ For $n=1$ ($b=1/3$), Eqns.~\atow-\valH\ reduce to
\eqn\atownone{a={4\over \left(1+2 \cos \left({2 \pi w_1\over 3}\right)\right)^2},\quad H= {2\over 9}{\left(1-2 \cos \left({2 \pi w_1\over 3}\right)\right)^2 \left(5+2\cos \left({2 \pi w_1\over 3}\right)\right)
\over \left(1+ 2 \cos \left({2 \pi w_1\over 3}\right)\right)^3}.}
We have then $(w_1)_{\rm max} =3/4$ and $a_{\rm max}=4$. Let us now show that the disappearance of a non-generic critical
solution beyond $a=a_{\rm max}$ may be associated to a {\it spontaneous domain symmetry} breaking in the model.
As already mentioned in Section 2.2, 
the $n=1$  fully-packed loop model with curvature weight $a$ is equivalent to an Ising model on random tetravalent maps
with an Ising coupling of the form $1+(a-1) \delta_{\varsigma \varsigma'}$ and a magnetic-like
coupling of the form $(h^{(1)})^2\delta_{\varsigma,1}+(h^{(2)})^2\delta_{\varsigma,-1}$ which, for $h^{(1)}=h^{(2)}=h$ reduces
to a weight $h^2$ per vertex of the tetravalent map. Up to a global factor $u^2$, we may also transfer the weight $u$ per green or
red vertex of the original triangulation, i.e.\ per face of the tetravalent map, to the vertices of the tetravalent map (since a
planar tetravalent map has two more faces than vertices), resulting in a total weight $h^2 u$ per vertex of the tetravalent map.

Now we may use the known solution for the above Ising model \BoulK\ (see also \BMS\ and \Bicubic): setting $H=(1-a^2)^2 h^2 u$ (in agreement with \valhsym\ at $n=1$),
the solution of the Ising model may be characterized by the equation (see for instance Eqn.~(6.4) of \Bicubic\ 
with the correspondence $g\leftrightarrow h^2 u$, $gz\leftrightarrow a$, $x\leftrightarrow H$)
\eqn\solIsing{P=1+3 H^2 P^3 +a^2 {P\over (1-3 H P)^2},}
where $(1-a^2)P$ is some appropriate generating function of the model (whose precise interpretation is
irrelevant to the discussion here). The locus of critical points is obtained 
by differentiating this equation with respect to $P$, hence by demanding that
\eqn\solIsingbis{1=9 H^2 P^2+a^2 {1+3 H P\over (1-3 HP)^3}.}
After factorization, this yields two branches of solutions: $a=(1-3 HP)^2$ and $(1+3 HP)=0$, corresponding respectively to a non-generic 
and a generic critical point. Plugging these values back in \solIsing\ and eliminating $P$ yields the following critical values
\eqn\criticIsing{\eqalign{H& ={2\over 9}(1-3 a + 2 a^{3/2}) \qquad\hbox{non-generic critical},\cr
H& ={3 a^2-8 \over 36}\qquad \qquad\qquad\   \hbox{generic critical}.\cr}}
The passage from the non-generic to the generic critical points occurs at the value $a=4$ (with a generic critical point
when $a>4$)
where the two determinations of $H$ are identical (note that these two determinations match up to order $(a-4)^3$ in the vicinity of $a=4$). In terms of spin variables, this passage corresponds precisely to an Ising transition, with a spontaneous breaking
of symmetry between the $+1$ and $-1$ spin domains for $a>4$.  

We see that the parametrization \atownone\ reproduces precisely the value of the non-generic critical point while
the value $a_{\rm max}=4$ corresponds precisely to the value of the Ising coupling at the Ising transition.
We may thus associate the disappearance 
of the non-generic solution for $a>a_{\rm max}$ with a spontaneous symmetry breaking between red and green domains
in the initially symmetric twofold fully-packed loop model. We expect that this conclusion, obtained for $n=1$ via the above equivalence,
also holds when $n\neq 1$. Heuristically, for a large enough curvature weight, the loops are forced to encircle small domain of one color in a background of the other color. Note that, for $n\neq 1$,  we have however no direct proof of the symmetry breaking.

\newsec{Conclusion}

In this paper, we have shown that the nested loop approach, initiated in \BBG, can be applied to more complicated loop models than those of the  $O(n)$-type, which incorporate a bending energy $a$ for the loops (as in \BBGa), and possibly display a domain symmetry breaking. With this method, one can obtain a simple and elegant combinatorial derivation of the functional relations satisfied by the generating functions of the model on planar maps, which may in some cases be solved analytically. This framework includes the $Q = n^2$ Potts model on general planar maps, which is equivalent to a particular fully-packed loop model.

We find that imposing an explicit domain symmetry breaking ($r \neq 1$) does not destroy the presence of non-generic critical points: the non-generic critical variety has codimension $2$ instead of $1$. More precisely in the fully-packed case, for each value of $(r,a)$ with $a < a_{\max}[r]$, there exists a non-generic critical point $(h_1,h_2)$. We underline that, whenever $r \neq 1$, $h^{(1)} \neq h^{(2)}$ so that the critical point is not equivalent to that of a dense $O(n)$ model, though we find that they lie in the same universality class. This conclusion applies to the critical Potts model on general random maps, which is not self-dual whenever $Q \neq 1$, and does not map to a critical $O(n)$ model. Even with fully-packed loops, we expect that one may reach the universality class of a dilute $O(n)$ model by simply tuning the curvature weight to $a = a_{\max}[r]$. We expect that $a_{\max}[r]$ is always strictly larger than $1$ and therefore the dilute phase cannot be reached strictly speaking in the Potts model (which assumes $a = 1$) without defects. For $a > a_{\max}[r]$, we expect a phase separation between regions with high concentration of red/low concentration of green, and vice versa. For the symmetric case ($r = 1$), this phase separation corresponds to a spontaneous breaking of symmetry between red and green domains. This is corroborated by known results for $n = 1$, as the model is then equivalent to a Ising model on tetravalent maps. 

Others statistical physics model on planar random maps, such as the 6-vertex model solved in \PZinnsixv-\Kostovsixv, the ADE models \KostovADE\ and generalizations thereof, might be treated with the same combinatorial formalism. The possible solution of those type of enumeration problems in all topologies by a ``topological recursion formula'' is currently under investigation \BEO.

\medskip

\noindent {\bf Acknowledgments.} We thank Nicolas Curien, Hugo Duminil-Copin and Paul Zinn-Justin for useful discussions.

\listrefs

\end